\documentclass[preprint,aps,floatfix]{revtex4}

\usepackage{amsmath}
\usepackage{graphicx}
\usepackage{amssymb}
\usepackage{epstopdf}
\usepackage{braket}
\usepackage{bm} %bold math

\usepackage{longtable}
\usepackage{multirow}
\usepackage{enumerate}

\usepackage{ragged2e}
\usepackage{lipsum}% for dummy text

\usepackage{rotating}
\usepackage{pdflscape}
\usepackage{afterpage}
\usepackage{capt-of}% or use the larger `caption` package
\usepackage{lipsum}% dummy text

\usepackage{hyperref}
\hypersetup{
    colorlinks=true,       % false: boxed links; true: colored links
    linkcolor=blue,        % color of internal links
    citecolor=blue,        % color of links to bibliography
    filecolor=blue,        % color of file links
    urlcolor=blue,         % color of external links
    runcolor=blue
}

\usepackage{xcolor}

\def \x {\mathbf{x}}

\begin{document}

\title{Molecular polaritons for controlling chemistry with quantum optics}

\author{Felipe Herrera}
%\email{felipe.herrera.u@usach.cl}
\affiliation{Department of Physics, Universidad de Santiago de Chile, Av. Ecuador 3493, Santiago, Chile.}
\affiliation{Millennium Institute for Research in Optics MIRO, Chile.}

\author{Jeffrey Owrutsky}
\affiliation{U.S. Naval Research Laboratory, Washington, USA}

\date{\today}            

\begin{abstract}
This is a tutorial-style introduction to the field of molecular polaritons. We describe the basic physical principles and consequences of strong light-matter coupling common to molecular ensembles embedded in UV-visible or infrared cavities. Using a microscopic quantum electrodynamics formulation, we discuss the competition between the collective cooperative dipolar response of a molecular ensemble and local dynamical processes that molecules typically undergo, including chemical reactions. We highlight some of the observable consequences of this competition between local and collective effects in linear transmission spectroscopy, including the formal equivalence between quantum mechanical theory and the classical transfer matrix method, under specific conditions of molecular density and indistinguishability. We also overview recent experimental and theoretical developments on strong and ultrastrong coupling with electronic and vibrational transitions, with a special focus on cavity-modified chemistry and infrared spectroscopy under vibrational strong coupling. We finally suggest several opportunities for further  studies  that may lead to novel applications in chemical and electromagnetic sensing, energy conversion, optoelectronics, quantum control and quantum technology. 

\end{abstract}

\maketitle

\section{Introduction}

Quantum optics traditionally concerns  the preparation of light having non-classical statistical properties \cite{Scully-book}, which is essentially a quantum control task. In cavity quantum electrodynamics (QED \cite{Mabuchi2002}), this goal can be achieved through the energy exchange between quantum emitters and a confined electromagnetic vacuum  \cite{Raimond2001}. In chemistry, quantum control of molecular and material processes using external fields has been a long-standing goal for decades \cite{Shapiro-Brumer-book,Tannor1988}. In contrast with quantum optics, where control schemes often involve temporal manipulation of a small number of variables \cite{Buluta2011}, coherent control of molecular processes typically require a targeted exploration of a multi-dimensional landscape of control parameters \cite{Dantus2004,Sussman2006}, even for  systems in isolation from their environments. Excitation dynamics in biological systems is a particularly complex problem for quantum control, given the large number of strongly-interacting electronic and nuclear degrees of freedom involved \cite{Cheng:2009}. Despite the complexity, quantum optimal control schemes \cite{Weinacht2002,Brif2010} can still be designed to steer the system dynamics towards a desired objective \cite{Caruso2012,Hoyer2014}.

Coherent control protocols in molecular systems often rely on perturbative linear or nonlinear light-matter interactions \cite{Shapiro-Brumer-book}. In order for the control lasers to imprint their amplitude and phase information onto a material wavefunction, it is best for the matter and field degrees of freedom to evolve independently.  Although perturbative coherent control schemes are often simpler to understand and design, strong field schemes for coherent control of population transfer \cite{Shore1991,Dudovich2005,Chen2010} are known to be more robust to protocol imperfections \cite{Vitanov2017}, and have enabled several results such as the formation of ultracold molecules \cite{Ospelkaus2008}, light-induced chemical dynamics \cite{Corrales2014}, single-photon transistors \cite{Gorniaczyk2014}, deterministic single-photon sources \cite{Wei2014}, and noise-resilient quantum gates \cite{Beterov2016,Yale2016,Kumar2016,Du2016,Yan2019}. 

Despite their conceptual differences, the fields involved in weak field and strong field quantum control protocols are largely classical and have large mean photon numbers. Therefore, the field amplitudes can be safely regarded as  scalar parameters that drive the evolution of a target material system. In this semiclassical regime, the quantum state of light does not undergo evolution, apart from trivial propagation effects. On the other hand, in quantum optics the photon number statistics of an incoming control field could in principle change due to free evolution of the coupled light-matter system, an effect that is not possible in the semiclassical picture of light-matter interaction so often used in  molecular spectroscopy \cite{Mukamel:2004}. Altering the quantum state of light would correspond to a type of {\it quantum back-action} on the control field that researchers can take into account when designing novel optimal control strategies with quantum light. The increased complexity and potential scope of such non-perturbative quantum optical control schemes could  open novel prospects for manipulating chemical reactions and material properties. 
 
In recent years, several experimental groups have used a diverse set of photonic structures to establish the possibility of manipulating intrinsic properties of molecules and molecular materials under conditions of strong and ultrastrong light-matter coupling with a confined electromagnetic vacuum in the optical \cite{Hutchison2012,Schartz2011,Orgiu2015,Ebbesen2016,Barachati2018,Daskalakis2017,Kena-Cohen:2008,Kena-Cohen:2010,Kena-Cohen2013,Lerario2017} and infrared \cite{Long2015,Kapon2017,Muallem2016,Saurabh2016,Simpkins2015,Thomas2016,Vergauwe2016,Chervy2018,George2015,George2016,Hertzog2017,Shalabney2015coherent,Shalabney2015raman,Dunkelberger2016,Xiang2018,Ahn2018,Dunkelberger2018,Thomas2019,Dunkelberger2019} regimes. This growing body of experimental results have positioned molecular cavity systems as novel implementations of cavity QED that complement other physical platforms with atomic gases \cite{RMiller2005}, quantum dots \cite{Reithmaier2004}, quantum wells \cite{Norris1994}, or superconducting circuits \cite{Blais2004}. Molecular cavities under strong and ultrastrong coupling lead to the dynamical formation of {\it molecular polaritons}: hybrid energy eigenstates  composed of entangled photonic, electronic and vibrational degrees of freedom. 
 
  The formation of molecular polaritons in optical and infrared cavities can offer viable routes for pursuing coherent control of molecular processes in condensed phase and room temperature, possibly without involving external laser fields but only vacuum effects. Moreover, these novel cavity systems may enable advances in the development of integrated photonic quantum technology \cite{OBrien2009}. Further opportunities for technological applications are expected to emerge from the study of light-matter interaction in exotic coupling regimes \cite{Forn-Diaz2018,Kockum2019}.

The growing literature on molecular polaritons has already been reviewed extensively. We refer the reader to previous reviews for a thorough description of previous literature. Most of the early experimental work on strong coupling with J-aggregates is reviewed in Refs. \cite{Tischler2007,Holmes2007}. Recent demonstrations of cavity-modified chemistry and molecular properties are reviewed in Ref. \cite{Ebbesen2016,Hertzog2019,Dovzhenko2018}. Strong light-matter coupling of molecular transitions near plasmonic nanoparticles is reviewed in Refs. \cite{Torma2015,Baranov2017,Cao2018,Wang2018}. For a description of the early work on organic polariton spectroscopy, we refer the reader to Refs.  \cite{Agranovich2005,Litinskaya2006,Michetti2008}. Recent molecular polariton theory is reviewed in Refs.  \cite{Herrera2017-review,Ribeiro2018a,Feist2018,Flick2018aa,Ruggenthaler2018}. 

This Perspective  is intended to serve as a tutorial-style introduction to the field of molecular polaritons. We first provide a brief account of pioneering experimental results, then discuss with relative detail the basic physical principles of intracavity light-matter coupling that are common to both optical and infrared cavities. We focus on the interplay between local and collective effects from a microscopic quantum description, an important topic that is a largely ignored in the literature. We finally discuss recent theoretical and experimental progress in the field, highlighting future challenges and opportunities for further research.

\section{Early experiments}
\label{sec:origins}

The term {\it polariton} refers to an excited energy eigenstate that describes the interaction of a quantized electromagnetic field with a material excitation. In crystalline solids, polaritons are quasi-particles with well-defined energy and momentum \cite{Agranovich-book}, although polaritons not only form in crystals. Exciton-polariton quasiparticles were introduced independently by Agranovich and Hopfield in the early 1960's to describe the microscopic origin of the dielectric constant of materials \cite{Hopfield1960,Agranovich1962}. Due to the maturity of semiconductor fabrication techniques, exciton-polaritons have been widely studied in inorganic semiconductor microstructures \cite{Skolnick1998,Vahala2003}, mostly at cryogenic temperatures. In 1997, Agranovich introduced the idea of using Frenkel excitons in organic semiconductors to enhance the emission properties of polaritons in dielectric microcavities \cite{Agranovich1997}. It was soon demonstrated by Lidzey that an anthracene crystal could reach the strong coupling regime with the vacuum field of a dielectric microcavity  at room temperature \cite{Lidzey1998}. This was a key innovation in comparison with inorganic semiconductors, as Wannier excitons are likely to undergo charge separation at room temperature because of their lower binding energies \cite{Agranovich-book}.  Wannier excitons form in crystalline inorganic semiconductors \cite{Vahala2003} and Frenkel excitons in organic materials \cite{Agranovich-book}. Among other fundamental studies and applications, strong light-matter coupling with dense films of organic chromophores and molecular aggregates in optical microcavities was initially explored as a route to enhance the emission properties of organic light-emitting devices \cite{Lidzey1999,Hobson2002,Kena-Cohen:2008} and lower the threshold for organic lasing \cite{Kena-Cohen:2010}. 

\begin{figure}[t]
\includegraphics[width=0.7\textwidth]{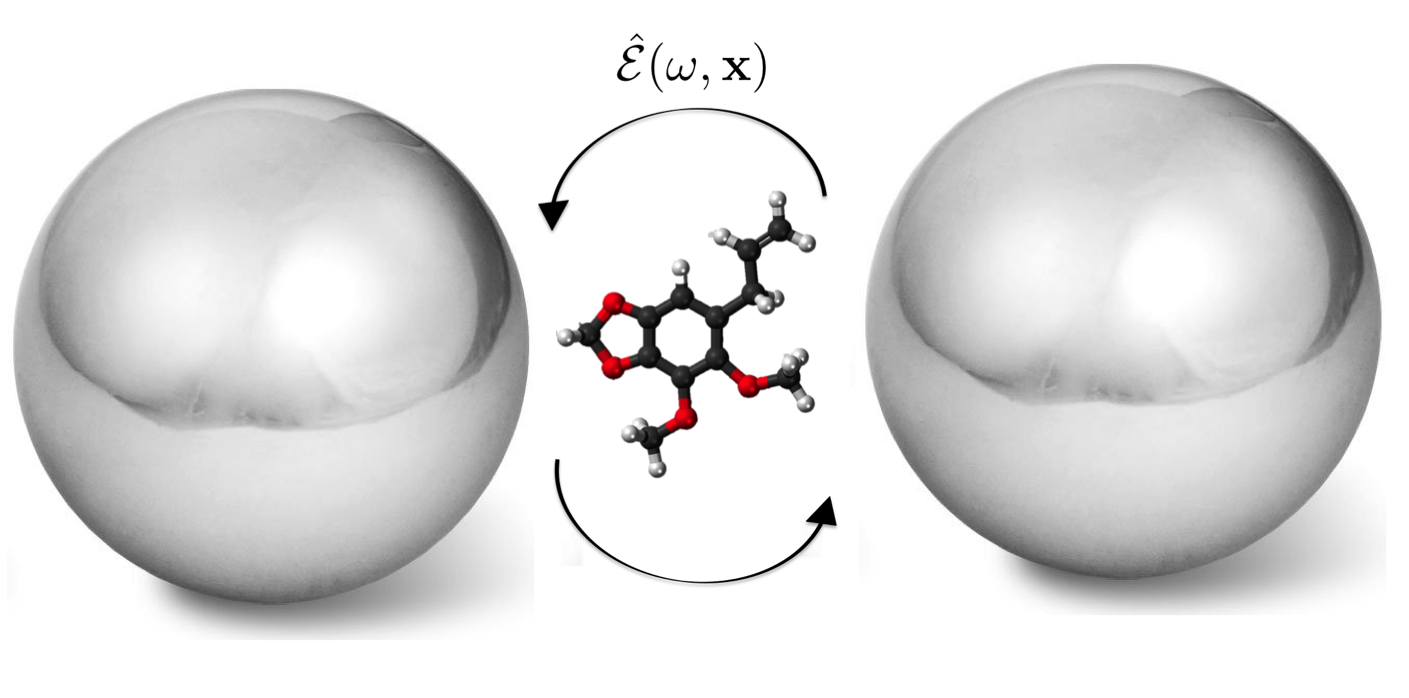}
\caption{Schematic nanoscale optical cavity made with two spherical nanoparticles with an individual organic chromophore located in the near field hot spot (gap). In the strong coupling regime, electronic or vibrational excitations of the molecule can coherently exchange energy with the dispersive and spatially inhomogeneous quantized near field $\hat {\mathcal{E}}(\omega,\mathbf{x})$. }
\label{fig:nanocavity}
\end{figure}

In the search for novel metal-based organic microcavities, it was later shown that room-temperature strong coupling between organic excitons and surface plasmon resonances could also be achieved \cite{Bellessa2004,Ditinger2005}. This extended the range of possible applications of plasmonic nanoparticles beyond enhanced sensing and spectroscopy \cite{Gonzalez2013,Cacciola2014,Bellessa2014,Delga2014}. More recently, quantum optical effects with a single  organic chromophore under strong electronic coupling in the near field have also  been demonstrated using plasmonic \cite{Chikkaraddy2016,Benz2016,Chikkaraddy2018} and dielectric  \cite{Wang2017} nanocavities. In Fig. \ref{fig:nanocavity} we illustrate a single-molecule cavity composed of metal nanoparticles.

After the pioneering demonstration by the group of Ebbesen \cite{Hutchison2012} of modified photoisomerization of merocyanine derivatives inside optical microcavities under electronic strong coupling, the field of molecular polaritons experienced a renewal of interest from a diverse group of experimental and theoretical researchers in chemistry, physics and materials science. In a few years the field has evolved into a challenging and fertile area in which researchers with complementary expertise must combine efforts in order to properly address open questions. It can be expected that relevant advances in quantum nanophotonics \cite{Tame2013}, microfluidics and nanofluidics \cite{Vanderpoorten2019}, electron microscopy \cite{Li2017},  spectroscopy with quantum light \cite{Dorfman2016,Raymer2013}, and quantum control theory \cite{Brif2010}, open exciting possibilities for the development of integrated quantum devices able to manipulate chemical systems by exploiting light-matter hybridization effects intrinsic to molecular polaritons.

\section{Molecular polariton basics}
\label{sec:basics}

Experiments have shown that any meaningful analysis of the spectroscopy and dynamics of molecular polaritons in the optical and infrared regimes must take into account several specific features of the molecular species and the photonic structures involved. These details include--but are not limited to--an accurate knowledge of the electronic and vibrational structure of the material system, the geometry and dielectric functions of the materials that compose the cavity structure,  knowledge of the predominant material and electromagnetic dissipation channels, and the presence or absence of external coherent (laser) and incoherent (thermal) energy sources. On the other hand, it has also been observed that while the rate of some molecular processes can be significantly affected upon resonant interaction with cavity fields \cite{Ebbesen2016,Hertzog2019}, the dynamics of other material processes occurs at rates that are basically indistinguishable from a cavity-free scenario \cite{Eizner2019}. 

By combining recent developments in quantum optics theory \cite{Buhmann2007} with state-of-the-art electronic structure methods, it should be possible to construct a fully {\it ab-initio} quantum theory of molecular polaritons that can simultaneously make precise testable predictions about the optical and chemical response of molecular ensembles in confined electromagnetic fields. However, even if such theoretical framework is eventually developed, researchers are still likely to continue interpreting their experimental data using the classical and semiclassical modeling tools that have proven useful for providing a good qualitative picture of optical and the infrared cavities under strong coupling. 

The more widely used modeling tools to rationalize experiments are the {\it transfer matrix} method \cite{Macleod2013}, and a simplified approach to model light-matter coupling that we can call the {\it fitting matrix} method. The former is a classical electrodynamics approach based on Maxwell's equations in continuous media, and the latter is a semiclassical model that closely resembles the treatment of coherent light-matter coupling in multi-level atomic gases \cite{Fleischhauer:2005}. 

In the rest of this section, we briefly describe the main ideas behind the transfer matrix and fitting matrix methods for modeling molecular polaritons in the optical and infrared regimes, highlighting their strengths and limitations. We then provide a tutorial-style discussion of a more complete microscopic cavity quantum electrodynamics approach to describe molecular polaritons, explicitly showing the conditions under which the microscopic model would provide the same level of information about the coupled cavity system as the transfer and fitting matrix approaches. We also discuss the interplay between collective and local effects in molecular ensembles, and how this interplay determines the properties of what are commonly known as the {\it dark states} of a coupled cavity.

\begin{figure*}[t]
\includegraphics[width=1.0\textwidth]{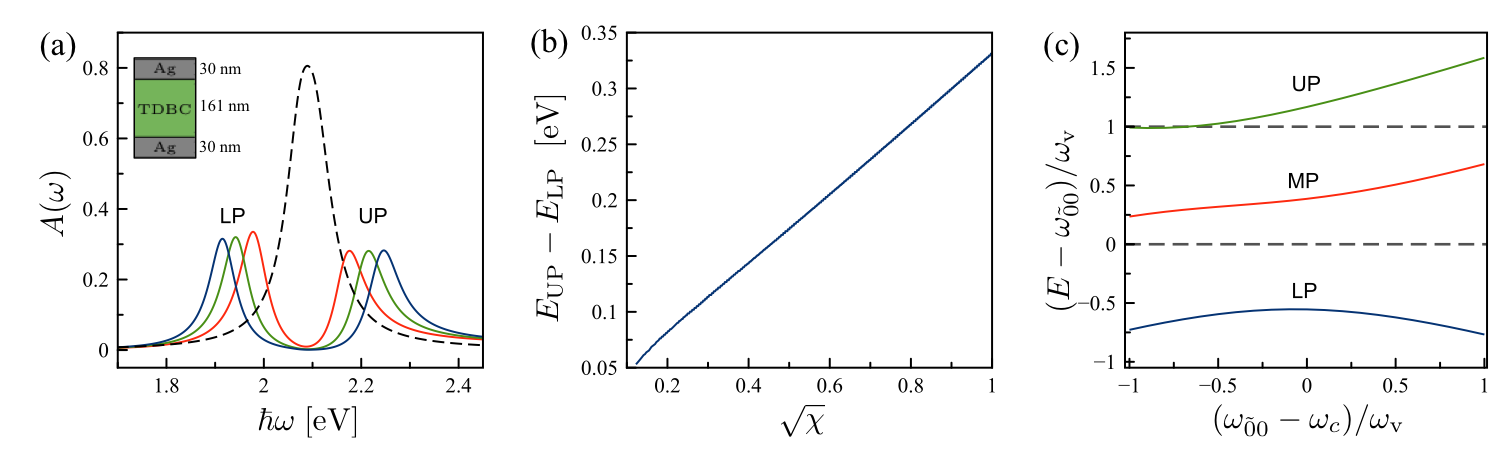}
\caption{(a) Absorption spectrum of a planar silver microcavity with TDBC J-aggregates (inset). The lower polariton (LP) and upper polariton (UP) peaks are shown for three values of the oscillator strength $\chi f_0$, with $\chi=1/3$ (red line), $\chi=2/3$ (green line), and $\chi=1$ (blue line). $f_0$ is the oscillator strength of a pure J-aggregate film with absorption spectrum (dashed line). Dielectric function parameters are taken from Ref. \cite{JohnsonChristy} for silver, and Ref. \cite{Zengin:2013} for TDBC aggregates. (b) Scaling of the UP-LP energy splitting as a function of $\sqrt{\chi}$. (c) Eigenvalues of a semiclassical fitting matrix model for an organic chromophore with two vibronic resonances, as a function of  detuning of the field frequency $\omega_c$ from the zero-phonon absorption peak. The resulting lower, middle and upper polariton (dressed) levels are shown. $\omega_{\rm v}$ is the vibrational frequency of the chromphore in the excited electronic state.}
\label{fig:general system}
\end{figure*}

\subsection{Classical transfer matrix method}

Consider the planar silver microcavity filled with a layer of TDBD molecular aggregates, as first used in Ref. \cite{Hobson2002} (Fig. \ref{fig:general system}a inset). The empty silver cavity is a Fabry-Perot resonator \cite{Macleod2013} with a resonant transmission peak near the absorption maximum of a bare TDBC film. Starting from Maxwell's equations with linear dielectric constitutive relations, together with the boundary conditions imposed by the multi-layer geometry, it is straightforward to derive relations between the incoming and outgoing electric and magnetic fields at each interface of this cavity nanostructure \cite{Macleod2013}. These input-output relations can be written in matrix form as
\begin{equation}
\mathbf{X}_b = \mathbf{M}(\omega) \mathbf{X}_a
\end{equation}
where $\mathbf{X}_a$ and $\mathbf{X}_b$ are two-component vectors that contain information about the electric and magnetic fields at two opposite silver-air cavity interfaces ($a$ and $b$), and $\mathbf{M}(\omega)$ is the so-called {\it transfer matrix} of the entire structure, which connects the incoming and outgoing fields at these border locations \cite{Macleod2013}. The frequency dependence of transfer matrix elements encodes the dispersive and absorptive properties of all the materials involved. In the linear regime, the optical response of materials is given by their dielectric function $\epsilon(\omega)$ \cite{Boyd-book,PALIK1997book}. The dielectric functions of each layer in the cavity are commonly parametrized using specific models (e.g., Drude-Lorentz \cite{Taflove2005}), with model parameters obtained by fitting spectroscopic observables. Once the dielectric functions and thicknesses of each layer are known, the transfer matrix $\mathbf{M}(\omega)$ can be used to compute the transmission ($T$), reflection ($R$) and absorption coefficients ($A\equiv 1-R-T$). Further adjustments of the dielectric function parameters for the active material may be needed to improve the fitting with cavity measurements \cite{Ebbesen2016}. 

In this classical optics approach, information about resonant light-matter coupling is obtained indirectly from splittings between peaks in a linear optical signal (transmission, reflection, absorption). No microscopic knowledge of the material system is needed to successfully interpret an experimental spectrum. In order to illustrate this point, consider the absorption spectrum shown in Fig. \ref{fig:general system}a. The dielectric constant of the TDBC layer is fitted to a Lorentz oscillator model to give \cite{Boyd-book,PALIK1997book}
\begin{equation}\label{eq:Lorentz model}
\epsilon(\omega) =\epsilon_\infty + \frac{f_0\omega_0^2}{\omega_0^2-\omega^2-i\gamma_0\omega},
\end{equation}
where $\epsilon_\infty$ is the dielectric background contribution, $\omega_0$ is the frequency of the zero-phonon (0-0) absorption peak of a bare TDBC sample, $\gamma_0$ is the associated resonance linewidth, and $f_0$ is the oscillator strength of the transition. The values of these parameters are obtained from experiments \cite{Zengin:2013}. The oscillator strength $f_0$ is in general another fitting parameter, although for a system of $N$ oscillating transition dipoles that are  {\it independent} (i.e., not correlated), $f_0$ can be shown to scale linearly with the number of dipoles using perturbation theory \cite{Boyd-book}.

Maxwell equations are such that the near resonant coupling of a cavity resonance with a single Lorentz absorption peak (dashed line in panel \ref{fig:general system}a), gives rise to two new normal modes (LP, UP) that increasingly split from each other as the value of $f_0$ grows. Let us rescale the oscillator strength in Eq. (\ref{eq:Lorentz model}) as $f_0\rightarrow \chi f_0$, with $0\leq \chi \leq 1$ being a free scaling parameter. Panel \ref{fig:general system}b shows that the splitting between LP and UP peaks then scales as $\chi^{1/2}$.

In experiments, the energy splitting between LP and UP peaks (Rabi splitting) is associated with the strength of light-matter coupling. If we {\it assume} that the free parameter $\chi$ is proportional to the number of emitters in the cavity, we then have a $\sqrt{N}$ scaling, commonly associated with collective light-matter interaction. However, we emphasize that {\it no quantum theory} was necessary to obtain the square-root scaling in panel \ref{fig:general system}b.  We later show that the square-root scaling in general does not hold for every molecular cavity system, but emerges from a microscopic quantum model for molecular transition dipoles in an ensemble that are indistinguishable.

The transfer matrix method is directly applicable to planar multi-layer structures \cite{Macleod2013}, and therefore has been widely used to analyze strong and ultrastrong coupling in optical microcavities and also infrared Fabry-Perot resonators \cite{Chantry1982}. In order to model light-matter coupling in plasmonic nanostructures using classical electrodynamics, a direct numerical or analytical solution of Maxwell equations is preferred \cite{Torma2015}. Numerical solvers such as the Finite-Difference Time Domain method (FDTD \cite{Taflove2005}) can be used to compute near fields and also simulate far field signals of  an active organic medium with known dielectric function that is in the near field of a plasmonic nanostructure with essentially arbitrary  geometry and material composition. Accurate analytical solutions for nanostructures without a well-defined symmetry have also been developed \cite{Li2016-plasmons,Li2018}. They provide valuable insight into the relative contributions of the different plasmon modes to light-matter interaction processes in experimentally relevant nanocavities. 

\subsection{Semiclassical fitting matrix method}

\begin{figure}[t]
\includegraphics[width=1\textwidth]{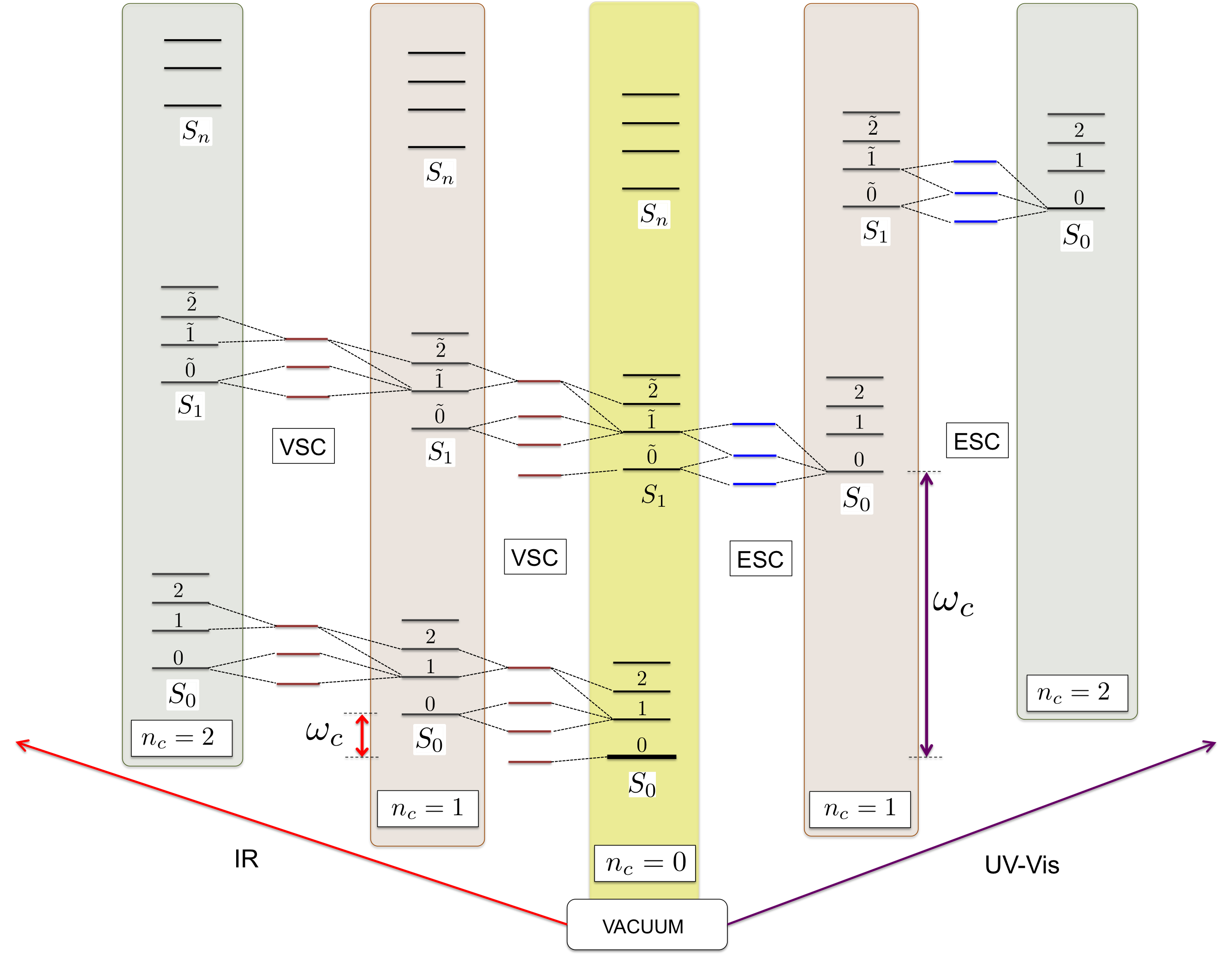}
\caption{Coupling scheme between diabatic molecule-photon states in the single-molecule regime. The center vacuum energy level diagram corresponds the bare electronic and vibrational energy level structure in vacuum ($n_c=0$). This includes vibrational manifolds in the two lowest electronic states ($S_0, S_1$), and multiple excited electronic manifolds $S_n$. To the right of the vacuum diagram we illustrate electronic strong coupling (ESC) with a single-mode cavity field at frequency $\omega_c$ in the UV-visible. For each cavity photon number $n_c\geq 1$, a diabatic replica of the entire electronic and vibrational spectrum can be defined, only shifted from the absolute ground level ($S_0, \nu=0, n_c=0$) by the energy $n_c\omega_c$. In the UV-visible, ESC between diabatic states with different photon number ($|\Delta n_c|=1$) leads to {\it vibronic polaritons} (blue energy levels). To the left side of the vacuum diagram, we illustrate vibrational strong coupling (VSC) for a cavity field frequency $\omega_c$ in the infrared. Coupling between diabatic states with different photon number ($|\Delta n_c|=1$) leads to the formation of {\it vibrational polaritons} (magenta energy levels) involving fundamentals and higher excited states (overtones) within the same electronic potential. }
\label{fig:diabatic couplings}
\end{figure}

The fitting matrix ($F$-matrix) method is another approach that has been used to interpret spectral signals of molecular polaritons in the UV-Vis \cite{Holmes2007}. It involves fitting the eigenvalues of a low-rank matrix to the frequencies of selected peaks obtained in cavity transmission, reflection or absorption. Since the quantum nature of the electromagnetic field is irrelevant to understand the linear  response of a strongly coupled cavity (see proof in Appendix \ref{app:Fmatrix derivation}), the $F$-matrix method can be derived from a semiclassical model. The corresponding Hamiltonian can be written as \cite{Scully-book}
\begin{equation}\label{eq:SC model}
\hat H(t) = \sum_{k}\omega_{k}\ket{E_k}\bra{E_k} +\sum_{k,k'\neq k} d_{kk'} \mathcal{E}(t)\ket{E_{k'}}\bra{E_k},
\end{equation}
where $\ket{E_k}$ describe the set of discrete energy eigenstates of the relevant molecular degrees of freedom (e.g., vibrational, vibronic), with eigenvalues  $\hbar\omega_k$. Dipole transitions between molecular eigenstates are determined by the dipole matrix elements $d_{kk'}$, in the presence of a classical monochromatic electromagnetic field $\mathcal{E}_0(t)=\mathcal{E}_0\cos(\omega_c t)$, with amplitude  $\mathcal{E}_0$ and frequency $\omega_c$. Equation (\ref{eq:SC model}) can be written in a rotating frame with respect to the external field \cite{Barnett-Radmore}, which eliminates the explicit time dependence of the light-matter coupling term. Several fundamental effects in coherent optics, including electromagnetically-induced transparency \cite{Fleischhauer:2005}, are determined by the eigenvalues and eigenstates of multi-level interaction Hamiltonians such as the one in Eq. (\ref{eq:SC model}). Semiclassical light-matter interaction eigenstates are known as {\it dressed states} \cite{Scully-book}. Manipulating the dynamics of dressed states has proven essential to develop several coherent population transfer techniques \cite{Vitanov2017}.    

We can illustrate the relation between the semiclassical model in Eq. (\ref{eq:SC model}) and the fitting matrix method with an example. Consider a single molecule embedded in an optical microcavity with cavity frequency $\omega_c$ in the UV-Vis, as illustrated in Fig. \ref{fig:diabatic couplings} (right side). We can describe vibronic transitions in the molecule using a displaced oscillator model \cite{Spano2010,Herrera2017-PRA}. The state $\ket{E_0}$ in Eq. (\ref{eq:SC model}) then corresponds to the lowest vibrational level ($\nu=0$) in the ground electronic state $S_0$. States $\ket{E_{\tilde 0}}$ and $\ket{E_{\tilde 1}}$ correspond to the lowest ($\tilde \nu=0$) and first excited ($\tilde \nu=1$) vibrational levels in the excited electronic state $S_1$, respectively. In free space (i.e., no cavity), the electric-dipole transitions $0\rightarrow\tilde 0$ and $0\rightarrow\tilde 1$ give rise to a vibronic progression in absorption from the absolute ground state $\ket{E_0}$ with transition frequencies $\omega_{\tilde \nu\nu}=\omega_{\tilde 00}+\tilde \nu \omega_{\rm v}$, where $\omega_{\tilde 00}$ is the zero-phonon transition frequency and $\omega_{\rm v}$ is the vibrational frequency in $S_1$. For this system, we can write Eq. (\ref{eq:SC model}) in a  frame rotating at the {\it fixed} classical frequency $\omega_c$, as a matrix given by
\begin{equation}\label{eq:Fmatrix}
\mathbf{{F}}({\omega_c})= \frac{1}{2}\left(\begin{array}{ccc}
-\Delta_{\tilde 00} & 2g_{\tilde 0 0}& 2g_{\tilde 1 0} \\
2g_{\tilde 0 0} &  \Delta_{\tilde 00}& 0 \\
2g_{\tilde 1 0} & 0& \Delta_{\tilde 00}+\omega_{\rm v}
\end{array}\right),
\end{equation}
where $\Delta_{\tilde 00}=\omega_{\tilde 00}-\omega_c$ is the detuning of the external field from the 0-$\tilde0$ transition frequency. The off-diagonal state-dependent Rabi frequencies are defined as $g_{\tilde \nu\nu}\equiv d_{\tilde \nu\nu}{\mathcal{E}}_0/2$. We show in Appendix \ref{app:Fmatrix derivation} that the  semiclassical matrix $\mathbf{F}(\omega_c)$ is formally equivalent to a truncated quantum model for a single-mode cavity field, which demonstrates that field quantization is {\it irrelevant} to describe linear transmission signals in strongly coupled cavities. 

In Fig. \ref{fig:general system}c, we plot the eigenvalues of the $F$-matrix in Eq. (\ref{eq:Fmatrix}) as a function of the detuning $\Delta_{\tilde 00}$. For concreteness, we  set $g_{\tilde 00}=\omega_{\rm v}$ and $g_{\tilde 10}=0.8\, g_{\tilde 00}$. The $F$-matrix gives rise to  lower, middle and upper dressed levels, which exhibit anti-crossing behavior near exact resonance ($\Delta_{\tilde 00}=0$). If we set $g_{\tilde 10}=0$ in Eq. (\ref{eq:Fmatrix}), the middle level disappears. The scaling of the energy difference between the resulting lower and upper dressed levels can be shown to scale linearly with $g_{\tilde 00}$ at exact resonance (not shown), which is the same behaviour in panel \ref{fig:general system}b for a single Lorentz resonance ($0\rightarrow\tilde 0$) in the transfer matrix method. 

In experiments, the values of the couplings $g_{\tilde \nu\nu}$ are not necessarily known from microscopic considerations. These coupling constants are usually inferred from a {\it fitting} procedure that compares the eigenvalues of the $F$-matrix at fixed $\omega_c$ with the positions of relevant peaks in cavity transmission, reflection or absorption measurements. The couplings could also be estimated using transfer matrix theory \cite{Simpkins2015}. The dispersive character of optical and infrared cavities can be taken into account in the fitting process by diagonalizing $\mathbf{F}(\omega_c)$ for different values of $\omega_c$ along the empty cavity dispersion curve. 

\subsection{Microscopic cavity QED approach}

Molecular polaritons are energy eigenstates of a light-matter interaction Hamiltonian that describes a cavity system in the optical or infrared regimes, with the electromagnetic field being a quantum mechanical operator. In other words, polaritons are the quantum analogues of the semiclassical dressed states discussed in the previous section. Unlike their semiclassical analogues, polariton eigenstates can be used describe the full quantum statistics of electromagnetic field observables \cite{Carmichael-book2}, as well as the dependence of the  field observables on the internal degrees of freedom of the molecular system \cite{Hernandez2019}. This type of microscopic understanding of molecular polaritons {\it is not} available in the transfer matrix or fitting matrix methods. In this section we discuss polariton systems from a microscopic quantum mechanical point of view.

Consider a general cavity system composed of $N$ inhomogeneously broadened molecular dipoles embedded in an optical or infrared cavity. We start from a inhomogeneous ensemble first and then discuss the conditions under which the commonly used homogeneous results emerge. We describe light-matter coupling in the point-dipole approximation within a multipolar framework \cite{Andrews2018}, to give the Hamiltonian
\begin{equation}\label{eq:generic Hamiltonian}
\hat{\mathcal{H}} = \hat H_{\rm c} + \sum_{i=1}^{N}\left(\hat H_{\rm m}(i)+ {\mathbf{d}}_i\cdot {\hat{\mathbf{D}}}(\x_i) \right).
\end{equation}
The first term describes an empty cavity (no molecules). This term is in general defined by the electromagnetic energy density over the optical structure, and takes into account the dispersive and absorptive character of the cavity materials \cite{Andrews2018}. The term $\hat H_{\rm m}(\x_i)$ describes the electronic, vibrational, and rotational degrees of freedom of the $i$-th molecule in the ensemble, located at position $\x_i$. We ignore direct electrostatic or retarded interaction between molecules. $\mathbf{d}_i$ is the electric dipole operator of molecule $i$, including electronic and nuclear charges, and $\hat{\mathbf{D}}(\x_i)$ is the dielectric displacement field operator evaluated at the location of each molecular emitter. The light-matter coupling model in Eq. (\ref{eq:generic Hamiltonian}) can in principle be used to interpret any experiment for which electric dipole coupling is relevant, provided that we accurately know each of the operators involved. This is currently unfeasible in general for molecular ensembles ($N\gg1$), but we show below that under a minimal set of assumptions, the general expression in Eq. (\ref{eq:generic Hamiltonian}) can be rewritten in a way that can be directly used to interpret experiments, and give a microscopic justification to the transfer matrix and fitting matrix approaches.

\subsubsection{Light-matter coupling in the homogeneous limit}

The system Hamiltonian in Eq. (\ref{eq:generic Hamiltonian}) is strictly local in the assumed point-dipole approximation. Light-matter coupling  can thus be considered as the quantized version of the semiclassical Autler-Townes effect \cite{Fleischhauer:2005}. In condensed phase, this local electric dipole interaction competes with other electrostatic shifts induced locally by molecules in the environment. For a molecular ensemble embedded in an optical or infrared cavity, a simple estimate shows that the light-matter interaction strength must be locally very weak, for typical values of the polariton splittings observed in experiments. This raises an interesting question: if light-matter coupling is locally weak, {\it where do the strong coupling effects observed in experiments come from}? 

Intuitively, macroscopic cooperative behavior of molecular dipoles should emerge when the molecules in an ensemble become indistinguishable. Indistinguishability leads to the delocalization of molecular dipoles into a giant collective dipole that strongly interacts with light. Therefore, any physical or chemical process that is intrinsically local would  tend to destroy the strong collective coupling of molecular dipoles with the electromagnetic field. We can thus expect that {\it every} molecular polariton signal obtained in experiments  results from the competition between these two opposing effects: the cooperative exchange of energy between molecular transitions with the cavity field, which leads to polariton formation, and local coherent and dissipative processes that occur at the level of individual molecules. {\it The  specific details of this competition between local and collective effects are what ultimately determine the observed chemical, transport and optical properties of molecular polaritons in the optical and infrared regimes}. 

In order to formalize this intuition, let us relax the locality constraints in Eq. (\ref{eq:generic Hamiltonian}). First, we neglect all the spatial derivatives of the dielectric displacement field, i.e., $\hat{\mathbf{D}}(\x)\approx \hat{\mathbf{D}}$. Next, assume that all dipole vectors in the ensemble $\mathbf{d}_i$ are either equally oriented or uniformly distributed with respect to the spatial orientation of the dielectric displacement field (both situations are  equivalent \cite{Litinskaya2006-disorder}). Finally, we ignore local shifts of the molecular energy levels (inhomogeneous broadening) and other coherent local effects in the bare molecular system. This allows us to make the replacement $\hat H_m(i)\rightarrow \hat H_m$. Under this conditions, Eq. (\ref{eq:generic Hamiltonian}) can be rewritten in the simpler form
\begin{equation}\label{eq:homogeneous limit}
\hat{\mathcal{H}}= \hat H_c + \hat{H}_N + \frac{\sqrt{N}\Omega}{2}\;\hat{d}_N \hat D
\end{equation}
where $\hat H_N=\sum_i\hat H_m$, $\hat d_N=\sum_i\hat d_i/\sqrt{N}$, and $\hat D$ is the dielectric field operator. The single-particle Rabi frequency $\Omega$ is proportional to the magnitude of $\mathbf{d}\cdot\hat{\mathbf{D}}$ for each molecular dipole,  assumed identical for all molecules in the ensemble. Given the indistinguishability of molecular emitters that our conditions impose, the energy eigenstates of the many-body {\it homogeneous} Hamiltonian in Eq. (\ref{eq:homogeneous limit}) are delocalized and  have a well-defined permutation symmetry under particle exchange. 

Consider a truncated state space defined by a ground state $\ket{g}$ and excited state $\ket{e}$, which may equally well represent the two lowest electronic singlet states ($S_0$, $S_1$) in a molecule, or the two lowest vibrational levels $\nu=0$ and $\nu=1$ in the lowest singlet potential ($S_0$). For an ensemble of $N$ molecules, we can define the totally-symmetric collective excited state
\begin{equation}\label{eq:bright electronic state}
\ket{\alpha_0} = \frac{1}{\sqrt{N}}\sum_i^N\, \ket{g_1,\ldots, e_i,\ldots,g_N},
\end{equation}
with $\hat H_N\ket{\alpha_0}=\omega_e\ket{\alpha_0}$, being $\omega_e$ the relevant UV-Vis or infrared transition frequency of a bare molecule. We also define the collective ground state $\ket{G}=\ket{g_1,g_2,\ldots,g_N}$ with $\hat H_N\ket{G}=\omega_g\ket{G}$, where $\omega_g$ is the energy reference. In the collective basis $\{\ket{G},\ket{\alpha_0} \}$, we can thus write  Eq. (\ref{eq:homogeneous limit}) as
\begin{equation}\label{eq:Htotal matrix}
\hat{\mathcal{H}}= \hat H_c + 
\left(\begin{array}{cc}
\omega_g & \sqrt{N}\Omega_{eg}\,\hat D/2\\
\sqrt{N}\Omega_{eg}\, \hat D/2 & \omega_e
\end{array}\right)+ \sum_{k\geq 2}\hat H_N^{(k)},
\end{equation}
where $\Omega_{eg}\equiv \bra{e}\tilde d\ket{g}\Omega$, where $\tilde d$ is a dimensionless dipole operator with no diagonal elements in the $\{\ket{g},\ket{e}\}$ basis, as is typical for material states with well-defined parity. The terms $\hat H_N^{(k)}$ with $k=\{2, 3,\ldots,N\}$ in Eq. (\ref{eq:Htotal matrix}) describe the contributions of states with two or more molecular excitations in the ensemble. 

Molecular cavity experiments are typically either carried out in the dark or involve weak laser or electrical pumping. In this so-called linear regime, the nonlinear contributions to the Hamiltonian ($k\geq 2$) are irrelevant and can be safely ignored. One can also consider that at most one cavity photon is present in the system within a cavity lifetime, and that the electromagnetic field can be reduced to a single cavity mode with discrete frequency $\omega_c$. Under these conditions,  Eq. (\ref{eq:Htotal matrix}) with $\omega_g\equiv0$ can be written in matrix form as
\begin{equation}\label{eq:Htotal RWA}
\mathbf{{H}}= \left(\begin{array}{cc}
\omega_c & \sqrt{N}\Omega_{eg}/2\\
\sqrt{N}\Omega_{eg}/2 & \omega_e
\end{array}\right) -i\left(\begin{array}{cc}
\kappa/2 & 0\\
0 & \gamma/2
\end{array}\right),
\end{equation}
with eigenvalues given by
\begin{equation}\label{eq:complex energies}
E_\pm = \frac{\omega_e+\omega_c}{2}-i\,\frac{\kappa+\gamma}{4}\pm \frac{1}{2}\sqrt{N\Omega_{eg}^2-\left(\frac{\kappa-\gamma}{2}-i\Delta_c\right)^2}, 
\end{equation}
where $\Delta_c=\omega_e-\omega_c$ is the  detuning of the single cavity mode from the molecular transition frequency.  The real parts of $E_-$ and $E_+$ are the energies of lower and upper polariton eigenstates, respectively.

The Hermitian contribution in Eq. (\ref{eq:Htotal RWA}) is formally equivalent to a two-level version of the semiclassical $F$-matrix, up to a trivial global energy shift (see derivation in Appendix \ref{app:Fmatrix derivation}).  The non-Hermitian contribution is introduced to account for cavity photon loss into the far field at rate $\kappa$, and coherence decay of the symmetric collective state at the rate $\gamma$, either via radiative relaxation, non-radiative relaxation, and pure dephasing. Radiative decay would in this case be superradiant relative to the bare molecular rate $\gamma_0$, i.e., $\gamma =\sqrt{N}\gamma_0$ \cite{Spano1989a}, due to the permutation symmetry of $\ket{\alpha_0}$. Therefore, the imaginary parts of $E_{\pm}$ correspond to the  homogeneous linewidths of the polariton peaks, as would be measured in a linear absorption experiment starting from the ground state $\ket{G,0_c}$.  Non-Hermitian Hamiltonians are used to describe dissipation of quantum systems in an approximate way. The approximation ignores the conditional evolution induced by quantum jumps on the system wavefunction \cite{Carmichael-book2}. 

The  polariton energies in Eq. (\ref{eq:complex energies}) exhibit an important feature that is often ignored. Even for a single-mode cavity that is on resonance with a molecular transition frequency, i.e., $\Delta_c=0$, the lower and upper polariton splitting in general cannot be expected to be symmetric around the molecular frequency $\omega_e$, {\it nor} scale as $\sqrt{N}$ with the molecule density. More explicitly, under resonant conditions we can write the polariton splitting as
\begin{equation}\label{eq:splitting}
\Delta E = 2\left[N\Omega_{eg}^2\left(1+\frac{1}{C_N}-\frac{1}{2C_N}\left(\frac{\gamma}{\kappa}+\frac{\kappa}{\gamma}\right)\right)\right]^{1/2},
\end{equation}
where
\begin{equation}\label{eq:cooperativity}
C_N = \frac{N\Omega_{eg}^2}{\gamma\kappa}
\end{equation}
is known as the collective cooperativity parameter \cite{Carmichael-book2,TANJISUZUKI2011}. Given that $\gamma\neq \kappa$ for most cavity systems, a necessary condition to observe a symmetric polariton splitting with a square-root-$N$ scaling would be $C_N> 1$. In other words, strong coupling emerges in the large collective cooperativity regime. 

We have thus shown that a microscopic many-particle Hamiltonian with local-only molecular contributions [Eq. (\ref{eq:generic Hamiltonian})] can support a collective basis under specific assumptions, in which the light-matter interaction Hamiltonian acquires a simple low-rank form [Eq. (\ref{eq:Htotal RWA})]. This corresponds to the homogeneous regime of light-matter interaction, where the cavity field cannot distinguish between molecules in the ensemble. In this homogeneous limit, the microscopic theory, truncated to the single-excitation manifold, is formally equivalent to the semiclassical $F$-matrix method described previously.  We also showed that taking into account the homogeneous linewidths of the material and cavity excitations ($\gamma$, $\kappa$), gives a polariton spectrum that coincides with the results of the $F$-matrix or transfer matrix methods in the large cooperativity limit $C_N\gg 1$. 

To further illustrate the equivalence between classical and quantum results of linear optical signals in the high cooperativity regime, compare Eq. (\ref{eq:splitting}) with the transfer matrix expression for the splitting in transmission for quantum well excitons in semiconductor microcavities in the low-reflectivity limit \cite{SAVONA1995}
\begin{equation}\label{eq:splitting QW}
\Delta E  = 2\sqrt{\sqrt{V^4+2V^2\gamma(\gamma+\kappa)}-\gamma^2},
\end{equation}
where $V=(2c\Gamma_0/(n_{\rm cav})L_{\rm eff})^{1/2}$ is the element of the transfer matrix that couples the exciton and cavity resonances (i.e., the classical analogue of $\Omega_{eg}$), $\gamma$ is the homogenous exciton linewidth, $\Gamma_0$ is the exciton radiative width, $\kappa=c(1-R)/(2n_{\rm cav}L_{\rm eff})$ is bare cavity linewidth, $R$ is the cavity reflectivity, $L_{\rm eff}$ is the effective cavity length,  $n_{\rm cav}$ the intracavity refractive index, and $c$ the speed of light. For large enough $V$, the transmission splitting in Eq. (\ref{eq:splitting QW}) becomes $\Delta E\propto \sqrt{\alpha_0\gamma/n_{\rm cav}}$, where $\alpha_0$ is the absorptivity of the medium. This relation has been used to accurately estimate the molecular concentration needed to observe a splitting the transmission spectrum of liquid-phase infrared cavities \cite{Simpkins2015}, i.e., the concentration required to reach the high cooperativity regime.

\subsubsection{The role of inhomogeneities}

Now  consider a more realistic scenario in which local shifts of molecular levels cannot be ignored. This is the usual case in solid and liquid-phase cavities, where vibrational and electronic absorption linewidths are dominated by inhomogeneous broadening \cite{Stuart2007,Spano2010}. In this case, the molecular transition frequency of the $i$-th molecular emitter can be written as
\begin{equation}\label{eq:energy disorder}
\omega_e(i) = \langle \omega_e\rangle +\Delta\omega_{i}
\end{equation}
where $\langle \omega_e\rangle$ would correspond to the center of an absorption band and $\Delta\omega_i$ is a static local frequency fluctuation (energy disorder), commonly assumed to be Gaussian with a standard  deviation $\sigma_E\ll \langle \omega_e\rangle$ \cite{Spano2010}. 

In a cavity, another type of static disorder corresponds to the unknown and possibly random location of emitters within the cavity field profile. The local Rabi frequency can thus be written as
\begin{equation}\label{eq:Rabi disorder}
\Omega_i =\langle \Omega\rangle + \Delta\Omega_i,
\end{equation}
where $\Delta\Omega_i$ is a local static fluctuation, to which we refer as {\it Rabi disorder}. The distribution of $\Delta\Omega_i$ is not well understood experimentally, as it depends on the cavity fabrication method. Efforts can be made to fill the cavity volume homogeneously with molecular dipoles such that $\Omega_i$ roughly matches the electric field profile of the cavity \cite{Baranov2017}. We ignore {\it orientational disorder} arising from the random orientation of molecular dipoles, as it has been shown that that the linear response of a random but uniformly orientated dipoles does not qualitative differ from a situation where all dipole are equally oriented \cite{Litinskaya2006-disorder}. This conclusion also holds for some coherent nonlinear optical signals \cite{Litinskaya2019}. 

Not all local effects in the material system are statically random. Molecules also have multiple sources of coherent and incoherent internal couplings between electronic, vibrational, rotational and spin degrees of freedom. These local couplings tend to destroy indistinguishability of molecules in an intracavity ensemble, since at any given time the internal state configuration of a specific molecule is in principle unknown and different from the rest. 

Consider for example vibronic coupling in organic chromophores. Intramolecular vibrations with frequencies $\omega_\xi$ can couple with an electronic transition $\ket{g_i}\leftrightarrow \ket{e_i}$ by dynamically modulating the electronic energy of the excited state \cite{May-Kuhn}. To lowest order in the mode displacements from equilibrium, this interaction leads to a local term of the form
\begin{equation}\label{eq:vibronic coupling}
\hat V_i = \sum_\xi \lambda_{\xi i} \,\omega_\xi (\hat b^\dagger_{\xi i} + \hat b_{\xi i})\otimes \ket{e_i}\bra{e_i},
\end{equation}
where $\hat b_{\xi i}$ is the mode operator for the $\xi$-th vibrational mode in the $i$-th molecule, and the dimensionless parameter $\lambda_{\xi i}$ quantifies the local electron-photon coupling strength via the spectral density $J_i(\omega)=\sum_\xi \lambda_{\xi i}^2\delta (\omega-\omega_\xi)$ \cite{May-Kuhn}. Due to vibronic coupling and the Condon principle, different molecules in an ensemble would have different probabilities of exchanging energy with a cavity vacuum, according to their instantaneous local vibrational configuration.

Current experiments often only probe global properties of a coupled molecular cavity system, although single-molecule local state addressing and manipulation may be possible in open cavities using nanotips .  Consequently,  any measured optical or chemical polariton signal must simultaneously carry information about collective homogeneous effects as well as local inhomogeneities. Therefore, the interpretation of observables based on homogeneous-only approaches such as the $F$-matrix method or its quantum analogue, should necessarily be incomplete. 

Ignoring inhomogeneities in a polariton model is equivalent to neglecting the role of collective material states that are not fully-symmetric with respect to permutations of molecular emitters. These non-symmetric collective states are commonly known as {\it dark states} of the cavity, akin to states without oscillator strength that emerge in molecular aggregates and molecular crystals \cite{Spano2010,Wurthner2011,Hestand2018}.

\subsubsection{How dark are {dark states} inside a cavity?}

To answer this question, begin by considering a translationally-invariant linear array of $N$ molecules in free space, with lattice constant $a$. For array excitations involving individual $S_0\rightarrow S_1$ transition dipoles, dark exciton states can be defined as \cite{Spano2010,May-Kuhn}
\begin{equation}\label{eq:dark exciton}
\ket{k\neq 0} = \frac{1}{\sqrt{N}}\sum_{i=1}^N{\rm e}^{i{k} {x}_i}\ket{g_1,\ldots,e_i,\ldots,g_N}, 
\end{equation}
where $x_i$ is the position of the $i$-th molecule in the array and ${k}$ is the exciton wavevector. In general, the exciton wavevector can take $N$  allowed values in the range $\{-\pi/a,-\pi/a +1,\ldots,0,\ldots,\pi/a \}$. For dark excitons, we must have $k\neq 0$, as only for the {\it bright} exciton we have $k=0$. The bright exciton is the only state that can formally couple with a resonant photon of wavelength $\lambda \gg a$ satisfying momentum conservation, leaving $N-1$ quasi-degenerate exciton states with $k\neq 0$ essentially uncoupled from the light-matter interaction process to first order due to momentum mismatch \cite{May-Kuhn}. The number of dark excitons grows linearly with $N$, therefore their density of states can easily exceed the bright exciton density in molecular aggregates and crystals. %Excitonic coherence in molecular aggregates and crystals is  due to near-field dipolar couplings or charge exchange between neighbouring molecules \cite{Spano}.

Despite they are formally uncoupled from light, dark exciton states ($k'\neq 0$) can still borrow oscillator strength from the bright exciton ($k=0$) due to vibronic coupling and static energy disorder \cite{Spano2010} via higher-order processes, and thus contribute to the aggregate absorption and emission spectra. In other words,  dark exciton states {\it are not} fully dark in realistic molecular aggregates, precisely because of local terms in the system Hamiltonian.

Consider now an ensemble of $N$ molecules inside an optical or infrared cavity. We assume that their individual Rabi frequencies are identical (i.e., no Rabi disorder), despite the fact that molecules may reside in unknown locations within the mode volume. We can describe single material excitations in a collective basis of the form
\begin{equation}\label{eq:alpha state}
\ket{\alpha} = \frac{1}{\sqrt{N}}\sum_{i=1}^{N}c_{\alpha i} \ket{g_1,\ldots,e_i,\ldots,g_N},
\end{equation}
where the vector $[c_{\alpha 1},c_{\alpha 2},\ldots, c_{\alpha N}]^{\rm T}$  physically encodes the relative phases for all  molecular dipoles in the ensemble. The totally-symmetric state in Eq. (\ref{eq:bright electronic state}) thus corresponds to the case where $c_{\alpha i}=1$ for $i=\{1,2,\ldots,N\}$. Spontaneous emission of a collective state $\ket{\alpha}$ occurs at a rate  $\gamma_\alpha = |\sum_i c_{\alpha i}\gamma_e|^2$, where $\gamma_e$ is the spontaneous emission rate of an individual molecule. Radiative decay of the collective state $\ket{\alpha}$ would thus be superradiant only for the totally-symmetric state, and there would be no radiative decay ($\gamma_\alpha=0$) for the remaining $N-1$ sets of $c_{\alpha i}$ coefficients that are not symmetric with respect to particle permutation. In other words, non-symmetric collective states are formally dark, despite the associated material excitations being {\it fully delocalized} over the  mode volume.  

The classification of collective material eigenstates into a bright manifold $\mathcal{P}=\{\ket{G},\ket{\alpha_0}\}$ and a dark manifold $\mathcal{Q}=\{\ket{\alpha_1},\ket{\alpha_2},\ldots, \ket{\alpha_{N-1}}\}$ is a trivial consequence of orthogonality of the $N$ vectors $[c_{\alpha 1},c_{\alpha 2},\ldots, c_{\alpha N}]^{\rm T}$, and has nothing to do with momentum conservation, as it is the case in molecular aggregates and molecular crystals. Comparison of Eqs. (\ref{eq:alpha state}) and (\ref{eq:dark exciton}) therefore suggests that a Frenkel exciton $\ket{k}$ in a translationally invariant system is only a special type of collective state $\ket{\alpha}$.  

%The symmetry of a collective state relative to permutation is encoded into transformation properties of the state coefficients $c_{\alpha i}$ under the exchange of molecular labels. It has been shown that by exploiting permutation symmetry it is possible to reduce the numerical complexity involved in simulating the dynamics of strongly-correlated atomic or molecular ensembles subject to driving and dissipation \cite{Gegg2017}. 

In was initially proposed by Litinskaya and Agranovich \cite{Litinskaya2006-disorder,Litinskaya2004}, that in a strongly coupled cavity, bright polariton states in which material excitations are delocalized over the entire cavity volume, could coexist and interact with a large density of incoherent dark excitons, over a common spectral range. For the set of intracavity dark excitations, the term {\it dark state reservoir} was coined. It is a reservoir in the sense that for large $N$, the number of dark states is overwhelmingly large in comparison with the number of bright polariton states. In the single-excitation manifold, the latter can be written on resonance as
\begin{equation}\label{eq:bright polaritons}
\ket{\Psi_\pm}= \frac{1}{\sqrt{2}}\left(\ket{\alpha_0}\ket{0_c}\pm \ket{G}\ket{1_c}\right),
\end{equation} 
where $\ket{0_c}$ and $\ket{1_c}$ represent the cavity vacuum and one-photon states, respectively. Just as in molecular aggregates it is possible for dark excitons $\ket{k'\neq 0}$  to borrow oscillator strength from a bright exciton $\ket{k=0}$ due to the presence of local inhomogeneities, {\it intracavity} dark states $\ket{\alpha\neq \alpha_0}$ can also borrow photonic character from a bright polariton state $\ket{\Psi_{\pm}}$ in the single excitation manifold, or beyond the one-excitation regime \cite{Herrera2017-PRA}. 

This type of {\it cavity photon borrowing} can occur due to {\it coherent} local effects in the polariton dynamics. For instance, inhomogeneities such as energy disorder, Rabi disorder and intramolecular vibronic coupling, can give rise to couplings between the totally-symmetric polariton states and non-symmetric material excitations \cite{Lopez2007,Spano2015,Herrera2016}. It is also known in quantum optics that local {\it dissipative} processes such as the presence of local non-radiative relaxation and dephasing, can also drive population away form a totally-symmetric polariton manifold into a dark state reservoir \cite{Shammah2017}. 
 Moreover, in molecular ensembles with strong vibronic coupling (e.g., rubrene), local effects can lead to the emergence of novel types of vibronic polaritons that have large photonic character, but give emission signals in regions of the spectrum that are seemingly dark in cavity absorption \cite{Herrera2017-PRL,Herrera2017-PRA}. 

\begin{sidewaystable}[h]
\caption{Recent experimental developments on vibrational strong coupling (VSC), vibrational ultrastrong coupling (VUSC), electronic strong coupling (ESC), and electronic ultrastrong coupling (EUSC).\label{tab:experiments}}
\begin{ruledtabular}
\begin{tabular}{c c c}
  Regime &      Description  &  Ref. \\\hline
ESC & Cavity-enhanced energy transfer and conductivity in organic media  &  \cite{Zhong2016,Zhong2017,Georgiou2018,Orgiu2015}\\
ESC/VSC & Strong coupling with biological light-harvesting systems  &  \cite{Coles2014-chlorosomes,Grant2018,Vergauwe2016,Vergauwe2019}\\
ESC & Cavity-modified photoisomerization and intersystem crossing &  \cite{Hutchison2012,Takahashi2019,Canaguier-Durand2013,Stranius2018,Eizner2019} \\
ESC & Strong coupling with an individual molecule in a plasmonic nanocavity & \cite{Chikkaraddy2016,Benz2016,Wang2017,Ojambati2019} \\
ESC & Polariton-enhanced organic light emitting devices &  \cite{Kena-Cohen:2010,Held2018,Barachati2018,Jayaprakash2019}\\
EUSC & Ultrastrong light-matter interaction with molecular ensembles  &\cite{Schartz2011,Held2018,Cacciola2014,Mazzeo2014,Gambino2015,Balci2013,Kena-Cohen2013}\\
VSC/VUSC &Vibrational polaritons in solid phase and liquid phase Fabry-Perot cavities& \cite{Shalabney2015coherent,Shalabney2015raman,Long2015,George2015,Muallem2016,Dunkelberger2016,George2016,Kapon2017,Xiang2018,Ahn2018,Dunkelberger2018,Chervy2018,Erwin2019}\\
VSC & Manipulation of chemical reactivity in the ground electronic state &\cite{Thomas2016,Thomas2019,Lather2019}
\end{tabular}
\end{ruledtabular}
\end{sidewaystable}

\section{Current status and  challenges }
\label{sec:challenges}

Research on molecular polaritons is thriving in terms of both experimental and theoretical developments. Even though experiments have been the driving force in the field, a fruitful interaction between theory and experiments is paving the way toward envisioned applications in quantum technology and quantum control. In this Section, we briefly overview selected recent theoretical and experimental results.

 \begin{sidewaystable}[t]
\caption{Recent theoretical developments on vibrational strong coupling (VSC), vibrational ultrastrong coupling (VUSC), and electronic strong coupling (ESC). \label{tab:theory}}
\begin{ruledtabular}
\begin{tabular}{c c c}
Regime& Description  &  Refs. \\
\hline
 ESC & Cavity-controlled intramolecular electron transfer in molecular ensembles. &  \cite{Herrera2016,Galego2016,Galego2017,Martinez-Martinez2018,Semenov2019}\\
VSC/ESC& Controlled chemical reactivity with spatially-separated donor and acceptor molecules.  &  \cite{Du2018-PARET,Du2019}\\
ESC & Cavity-enhanced energy and charge transport with molecular ensembles.  &  \cite{Feist2015,Schachenmayer2015,Hagenmuller2017,Hegenmuller2018}\\ 
ESC & Absorption and photoluminescence of vibronic polaritons in molecular ensembles.  &  \cite{Litinskaya2004,Michetti2008,Mazza2009,Cwik2016,Herrera2017-PRA}\\
VSC & Linear and nonlinear spectroscopy of vibrational polaritons in molecular ensembles. &  \cite{DelPino2015,delPino2015raman,Strashko2016,Ribeiro2018}\\
ESC & Polariton condensation and lasing with vibronic transitions in molecular ensembles.     &  \cite{Cwik2014,Strashko2018}\\
ESC &Few-photon nonlinear quantum optics with molecular ensembles. &  \cite{Herrera2014,Litinskaya2019}\\
 VUSC & Ultrastrong light-matter interaction with molecular vibrations  &  \cite{Hernandez2019,Flick2017}\\
ESC& Non-adiabatic electron-photon-nuclear dynamics using {\it ab-initio} and semiclassical methods.   &  \cite{Vendrell2018,Galego2015,Luk2017,Flick2017,Flick2017a,Flick2018,Flick2018a,Vendrell2018aa,Fregoni2018,Galego2019,Rivera2019,Triana2018,Triana2019,Rossi2019}\\
ESC& Driven-dissipative polariton dynamics using open quantum system methods.  &  \cite{delPino2018,Varguet2019,Reitz2019}\\  
\end{tabular}
\end{ruledtabular}
\end{sidewaystable}

\subsection{Recent experimental progress}

The field of molecular polaritons has been primarily driven by experimental breakthroughs. Several pioneering results have advanced the field in directions that may eventually lead to the development of room-temperature integrated quantum technology. Among these achievements we highlight the demonstration of strong coupling of organic molecules with optical microcavities \cite{Lidzey1998} and surface plasmons \cite{Bellessa2014}, ultrastrong coupling with organic photoswitches \cite{Schartz2011}, and the demonstration of vibrational strong and ultrastrong coupling in solid phase \cite{Long2015,Shalabney2015coherent} and liquid phase Fabry-Perot cavities \cite{George2015,Dunkelberger2016,George2016}. Many experimental and theoretical works directly build on these results. In Table \ref{tab:experiments}, we list some of the experimental problems explored so far. For reviews of earlier experiments, we refer the reader to Refs. \cite{Tischler2007,Holmes2007,Torma2015}.

 The demonstration by Hutchison {\it et al.} \cite{Hutchison2012} of a modified photoisomerization reaction rate under conditions of strong coupling with a cavity vacuum for one of the involved isomers, may be regarded as the origin of a new research field in physical chemistry \cite{Ribeiro2018a,Feist2018,Dovzhenko2018,Hertzog2019}.  There are several widely-used technologies that ultimately base their efficiency on the rates of chemical reactions or electron transfer processes that occur in excited electronic states (e.g., sunscreens, polymers, catalysis, solar cells, OLEDs). Therefore, the ability to manipulate the rates and branching ratios of these fundamental chemical processes in a reversible manner using light-matter interaction with a vacuum field, suggests a promising route for targeted control of excited state reactivity, without exposing fragile molecular species or materials to the damaging effects of intense laser fields. 
 
 Promising results in this direction are the  demonstrations of cavity-modified intramolecular electron transfer   between different electronic manifolds (singlet fission), under conditions of strong coupling of singlet states inside an optical microcavity \cite{Takahashi2019,Stranius2018,Eizner2019}. The proposed mechanism for this process involves the manipulation of the relative energy levels between electron donor and acceptor manifolds via polariton formation \cite{Herrera2016,Martinez-Martinez2018}. The reversible control of  singlet-to-triplet conversion may serve to improve the external quantum efficiency of organic optoelectronic devices.  
 
 While experiments on strong coupling with electronic transitions date back decades \cite{Lidzey1998}, similar studies for vibrational strong coupling (VSC) are more recent. Current experimental focus is on the  manipulation of chemical reactions in the ground electronic state inside Fabry-Perot cavities \cite{Thomas2016,Thomas2019,Lather2019}, and also cavity-controlled steady-state \cite{George2015,Shalabney2015coherent,Long2015,Simpkins2015,George2016,Vergauwe2016,Muallem2016,Kapon2017,Ahn2018,Imran2019,Erwin2019}  and ultrafast \cite{Dunkelberger2016,Xiang2018,Dunkelberger2018,Xiang2019manipulating,Xiang2019state,Dunkelberger2019} vibrational spectroscopy.  Chemical reactivity experiments under VSC are typically carried out in the dark, i.e., the cavity is not driven by external laser light, and thus the role of dark states may be different and potentially more important than the bright states, but in this not well understood. For example, modified reaction rates have been ascribed to dramatic changes (near 1 eV) in the activation barrier energies \cite{Thomas2019}, which are difficult to reconcile with the relatively small Rabi splittings on the order of 10  meV realized in infrared transmission. 
 
Infrared spectroscopy studies of VSC systems may shed light on the mechanism of cavity-modified chemistry. Initial spectroscopic studies on VSC were carried out on first-order Fabry-Perot cavities that were several microns long, involving carbonyl absorption bands of polyvinyl acetate \cite{Shalabney2015coherent} and polymethymethacrylate \cite{Long2015,Muallem2016}. VSC was quickly extended to  several neat liquids (diphenyl phosphoryl azide, benzonitrile) and to longer cavities \cite{George2015}.  It was important to achieve strong coupling in longer cavities to conveniently incorporate a wider range of chemically relevant materials, including species dissolved in solution as demonstrated for W(CO)$_6$ in hexane \cite{Simpkins2015}. In this sense, it is advantageous that the Rabi splitting scales with the concentration and not the cavity length (provided the concentration is preserved), so that longer cavities can be used without sacrificing coupling strength. Vibrational ultrastrong coupling (VUSC), in which the splitting is a substantial fraction of the vibrational frequency $\Omega/\omega_{0}>0.1$, was achieved with neat liquids of strong absorbers including CS$_2$ \cite{George2016}. Steady-state spectroscopy has also been used to show simultaneous coupling between a cavity mode and multiple vibrational absorptions in complex molecules and mixtures \cite{Muallem2016,Imran2019}.

Understanding the dynamics of vibrational polaritons will likely be key to understanding reactivity, especially for photo-initiated processes. To that end, time resolved, IR pump-probe and 2DIR studies of cavity-coupled W(CO)$_6$ in hexane have been used to identify transient effects in strongly coupled molecular ensembles. Initial pump-probe results showed two predominant features after excitation that transfered population from the ground vibrational state to excited states:  polariton contraction resulting from depletion of the ground state and a large contribution from dark excited state absorption. The dark state transient population was found to decay at the same rate as the excited vibration in free space, consistent with these dark states behaving much like free uncoupled molecules. In addition, features that relax more rapidly were tentatively attributed to polariton population changes \cite{Dunkelberger2016}. 

Subsequent studies with 2DIR have further elucidated the dynamics of vibrational polaritons in W(CO)$_6$ \cite{Xiang2018}. Because 2DIR studies give excitation-frequency resolution, they conclusively show that the polariton modes, and the dark reservoir, can be readily excited by incident light. Multidimensional spectroscopy also gives facile experimental access to excitation pathways that revealed coherent exchange between the upper and lower vibrational polaritons \cite{Xiang2019manipulating}. The coherence led to strongly modulated transmission across the polariton spectrum, with potential implications for VSC-based photonic devices. Extending 2DIR studies to W(CO)$_6$ in a series of organic solvents has recently shown that the homogeneous linewidth of the vibrational absorber has a strong impact on the observed dynamics \cite{Xiang2019state}.

Additional ultrafast studies with W(CO)$_6$ have explored the boundary between vibrational weak and strong coupling. Reducing the Rabi splitting in this material system reduces the overlap between the lower polariton and higher-lying states of the uncoupled vibration and simplifies the experimental spectra. More importantly, it was reported that these intermediate-coupling samples are excellent candidates for photonic devices. The transmission, reflection and absorption spectra are highly dependent on the incident power because the material absorption is easily saturated. As the absorption saturates, the Rabi splitting decreases. At sufficiently low excitation fluences, the spectrum can collapse to a single transmission peak at the empty-cavity mode frequency reversibly over sub-nanosecond timescales. 
 
 The time-resolved studies described above have, thus far, been limited to an ideal absorber and are not easily connected to the promising results reported for VSC-modified reactions. Under conditions of vibrational strong coupling in liquid phase, ground state chemical reactions can proceed through novel pathways that have yet to be fully characterized from a mechanistic perspective.  For some molecular species, reactions are inhibited and for other systems they are catalyzed. For instance, strong coupling of a Si-C bond in a substituted acetylene molecule with an infrared cavity vacuum led to a decreased rate of the Si-C bond breakage by a moderate factor in comparison with a control \cite{Thomas2016}. It has also been shown that  the solvolysis of {\it para}-nitrophenyl acetate (reactant) in ethyl acetate (solvent) proceeds up to an order of magnitude faster when the C=O bonds of both reactant and solvent are resonant with a cavity mode, but only the solvent is strongly coupled \cite{Lather2019}. In other words, by strongly coupling a molecular species in excess (solvent), it is possible to modify the chemical reactivity of a dilute component in solution, for which light-matter coupling is weak.  Despite recent theoretical efforts \cite{Hernandez2019,Campos-Gonzalez-Angulo:2019aa}, the underlying microscopic mechanism for these remarkable observations is not well understood. 

% On the other hand, the reaction rate for the hydrolysis of cyanate ions and ammonia borane under strong coupling with the broad OH infrared band water (solvent), has been preliminary reported to increase by up to four orders of magnitude (catalysis) in comparison with free space \cite{Hiura2018}. 

\subsection{Recent theoretical progress}

In Table \ref{tab:theory}, we list some of the recent theoretical research directions explored to date. The initial condensed-matter approach to organic polaritons \cite{Litinskaya2004,Litinskaya2006,Agranovich2005,Fontanesi2009,Mazza2009,Michetti2008}, has been gradually replaced in recent years by what can be called a ``quantum optics'' approach. The traditional atomic cavity QED theory \cite{Mabuchi2002} is here extended to take into account the relevant physics and chemistry of molecular degrees of freedom. This quantum optics approach was initially used by Cwik {\it et al.} \cite{Cwik2014} to describe polariton condensation in microcavities, and Herrera {\it et al.} \cite{Herrera2014} to describe light-by-light switching with molecular aggregates in  microcavities. The works listed in Table \ref{tab:theory} predominantly use a quantum optics approach to predict or interpret experimental observables. Among these results, the development of the Holstein-Tavis-Cummings (HTC) model of vibronic polaritons \cite{Herrera2016,Herrera2017-review} has been particularly useful, as it has been successfully used to explain features in the optical signals of organic microcavities that were notoriously confusing, such as the apparent breakdown of reciprocity in the oscillator strengths in absorption and emission strengths over specific frequencies \cite{Hobson2002,George2015-farad}. Further predictions of the HTC model regarding the nature of the lower polariton manifold were later confirmed using variational techniques \cite{Wu2016,Zeb2018}, and applied in subsequent work on singlet fission \cite{Martinez-Martinez2018a} and long-range energy transfer \cite{Du2018-PARET}. Recent experiments with rubrene microcavities \cite{Takahashi2019} confirm the validity of the HTC model to describe the optical response and chemical reactivity of molecular polaritons in the optical regime, thus consolidating the quantum optics approach to describe these systems. 

Another promising trend in the field is development of (semi) {\it ab-initio} methods for non-adiabatic molecular polariton dynamics, as first proposed by Feist {\it et al.}  \cite{Galego2015}. This integrative approach has been further developed by several groups \cite{kowalewski2016cavity,Flick2017,Luk2017,Flick2018,Schafer2018,Fregoni2018,Triana2018,Vendrell2018,Rivera2019}. The main strength of this methodology is its compatibility with conventional electronic structure and molecular dynamics techniques. In this approach, a single-mode cavity field is essentially treated as another {\it effective} nuclear degree of freedom to which electrons can strongly couple. In this coordinate-only formalism, polaritonic energy surfaces and reaction coordinates can be defined \cite{Galego2015,Galego2019}. This  methodology has been primarily focused on strong light-matter coupling with individual molecules, but  extensions to molecular ensembles are possible \cite{Vendrell2018aa,Groenhof2019}. Predictions made with this methodology however have yet to be confirmed in experiments.  

\begin{figure}[t]
\includegraphics[width=0.8\textwidth]{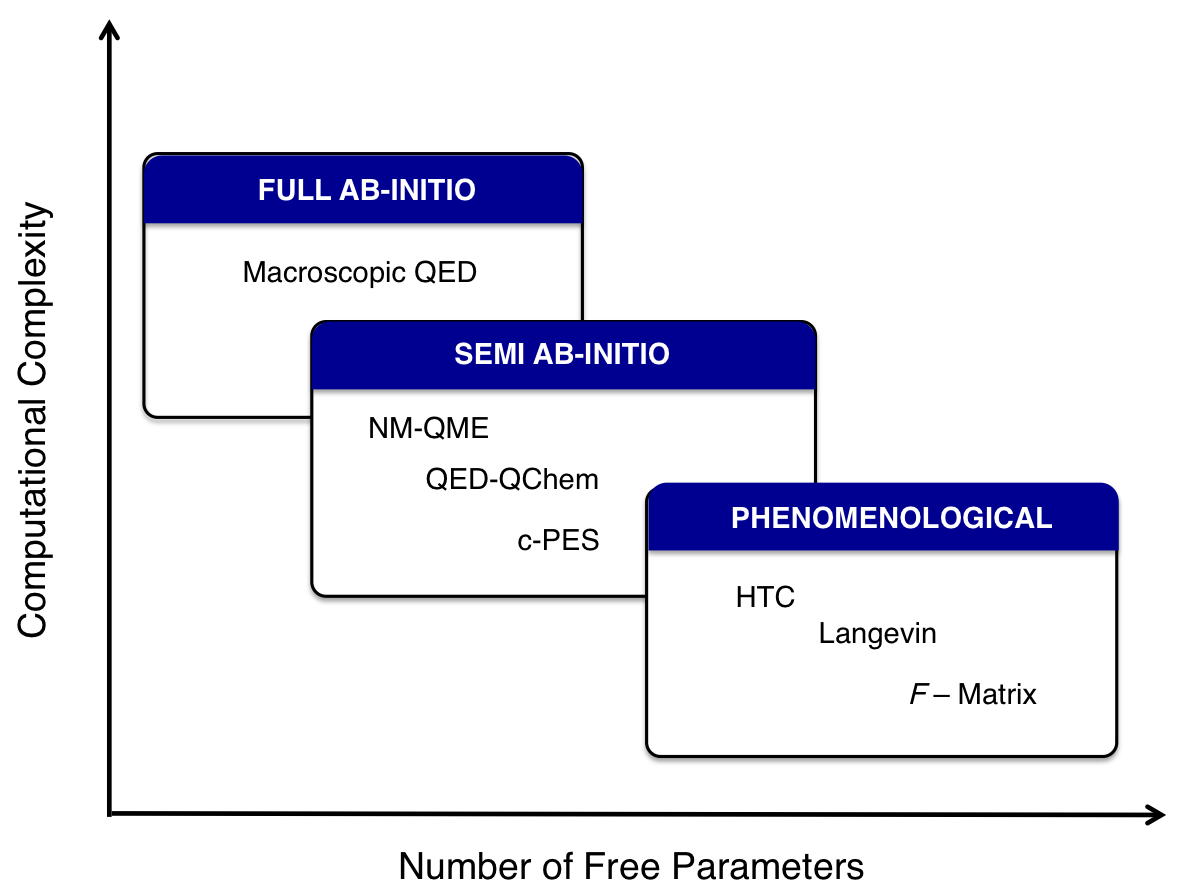}
\caption{Diagram of quantum mechanical methods for molecular polaritons in terms of the number of free parameters and numerical complexity. Fully phenomenological models describe both material and field degrees of freedom using semi-empirical Hamiltonians with parameters obtained from experiments. Semi {\it ab-initio} methods treat the material degrees of freedom using quantum chemical methods but the cavity field and the light-matter coupling parameters are obtained from experiments. Full {\it ab-initio} methods would also involve a description of the dispersive and absorptive properties of the cavity field using macroscopic quantum electrodynamics, which essentially has no fitting parameters. Classical transfer matrix theory cannot provide more information than phenomenological quantum methods. }
\label{fig:theory hierarchy}
\end{figure}

For classification purposes, we propose in Fig. \ref{fig:theory hierarchy} a complexity hierarchy for theory methods in the field. At the lower-right end of the complexity scale, we have the simple  phenomenological models, such as the semiclassical fitting matrix method ($F$-matrix), which involve a relatively large number of fitting parameters. At the upper-left end of the complexity scale, we have a  hypothetical theory of molecular polaritons where the electron and nuclear dynamics of molecules in an ensemble are treated using {\it ab-initio} methods with quantum chemical accuracy, and {\it also} the electromagnetic field is treated using macroscopic quantum electrodynamics \cite{Buhmann2007,Knoll2001,Raabe2007}, which generalizes empty-space quantum optics to treat dispersive and absorptive media. This superior fully {\it ab-initio} theory would have essentially no fitting parameters, and has yet to be developed. All other methods used to date, including the HTC model \cite{Herrera2016},  cavity potential energy surfaces (c-PES \cite{Galego2015}), QED quantum chemistry (QED-QChem \cite{Schafer2018}), non-Markovian quantum master equation (NM-QME \cite{delPino2018}) and Langevin equations \cite{Reitz2019}, would fit somewhere in between these two complexity extremes.

\section{Conclusion and Outlook}

Experimental efforts in the field of molecular polaritons have significantly evolved in two decades since the first demonstration of strong light-matter coupling with molecular crystals \cite{Lidzey1998}. Over the first decade,  the main focus  was the development of novel organic solid-state light emitting devices with polariton-enhanced  efficiencies \cite{Tischler2007}. Although this continues to be an important research direction that may lead to novel applications in quantum technology \cite{Kena-Cohen2010}, the idea of building infrared and optical cavity structures for controlling chemical reactions has gained significant traction in recent years \cite{Ebbesen2016}. Understanding the microscopic mechanisms that determine the rate of reactive processes in solid and liquid phases, for example may stimulate the development of novel coherent control techniques that exploit the quantization of the electromagnetic field to achieve a target reaction branching ratio. Moreover, the micrometer-size cavities in the infrared regime enable an integration with conventional microfluidics technology, which  would allow the study and control of intracavity chemical processes in solution. The use of mirror materials that are semi-transparent in the UV-visible could also enable the {\it in-situ} ultrafast optical monitoring of reaction intermediates.  Applications in polariton-assisted chemical and electromagnetic sensing can also be envisioned.

New research directions can emerge from a more focused theoretical and experimental study of coherent and dissipative dynamical effects that have been observed in molecular cavity systems. For example, our understanding of polariton lifetimes in optical microcavities is still far from complete \cite{George2015-farad}, possibly due to the difficulty of spectroscopically resolving the pathways for population transfer between molecular polaritons and the collective dark state manifold. Moreover, the interaction of molecular polaritons with their thermal reservoir under electronic strong coupling has only recently attracted experimental and theoretical interest \cite{Wersall2019,Semenov2019}.

 The  development  of technology based on the dynamics of molecular polaritons would thus benefit from a more active interplay between theoretical and experimental research. The recent experimental verification of the Holstein-Tavis-Cummings \cite{Herrera2016} using rubrene microcavities \cite{Takahashi2019} is a promising step in this direction. Experiments can also motivate the development of theoretical tools and protocols for cavity-controlled chemistry. For example, a recent proposal for tuning the rate of energy transfer between spatially separated donor  and acceptor molecular species \cite{Du2019} builds on the experimental demonstration of ultrafast manipulation of the  light-matter interaction strength in UV-visible \cite{Zhong2016,Zhong2017,Georgiou2018} and infrared \cite{Dunkelberger2018} cavities. This proposal for ``remote control chemistry" \cite{Du2019} may in turn stimulate the fabrication of unconventional infrared cavity structures, and the development of novel detection schemes. In the optical regime, detailed electronic structure calculations including polariton formation have recently predicted that the hydrolysis of tert-butyl chloride can be ``self-catalyzed" at the single-molecule level, in the near field of plasmonic nanoparticles without external stimuli \cite{Climent2019}. Testing this prediction would require the use of advanced single-molecule detection techniques, possibly in integration with nanofluidics  \cite{Bahsoun2018nanofluidics}. 
 
 Another problem where a collaboration between theory and experiments can be fruitful is strong coupling in the gas phase. Detailed theoretical predictions about spontaneous generation of infrared light with diatomic molecules under conditions of electronic strong coupling \cite{Triana2018,Triana2019} represent a challenge for experimental verification, as achieving the strong coupling regime with gas-phase cavities is comparably difficult, although in principle possible. This should present new challenges and opportunities given the additional degrees of freedom involved and prospects for state-to-state chemical studies, in particular at ultracold temperatures \cite{PerezRios2017}. Gas-phase strong coupling has already been implemented with atomic gases in photonic crystal cavities \cite{Ruddell:17}.

Strong coupling to vibrational transitions in gas-phase molecules in infrared cavities should also be considered. Even though the number density of species is much lower in gases than in condensed phase, the linewidths are considerably narrower.  Rotationally-resolved vibrational transitions of small diatomic or polyatomic molecules are attractive for strong coupling since they afford the opportunity to explore whether state to state processes including energy transfer and reactions can be modified and perhaps offer better opportunities to understand mechanistic details. Assessing the feasibility of this would require careful analysis of broadening processes in various pressure regimes. At low pressure (and higher frequency), rovibrational lines tend to be inhomogeneously broadened due to Doppler shifts. On the other hand, at higher pressures spectral lines are  homogeneously pressure broadened. For a pure gas  the peak absorption intensity remains constants and the line broadens as the pressure increases. Simple estimates suggest that in the pressure broadened regime, the condition for strong coupling reduces to the absorption cross section being larger than the pressure broadening coefficient. In addition, since increasing the pressure of the absorbing gas increases the Rabi frequency at the same rate as the linewidth, a cavity width that is matched to the absorption linewidth of the gas medium, would still lead to strong coupling. It might also be possible to reduce Doppler broadening in a jet or a trap configuration, so that strong coupling could be achieved at very low number densities.

Further research opportunities can arise from the integration of molecular polariton systems with traditional photonic and chemical technologies. For example, by using a nematic liquid crystal embedded in an infrared cavity with conductive mirrors \cite{Hertzog2017}, researchers were able to manipulate the light-matter interaction strength by applying an external voltage. The voltage was used to induce a macroscopic rotation of the medium polarization relative to the cavity field orientation. Further explorations in this direction may lead to the development of novel screen displays. Moreover, upon integration of liquid crystal cavities with microfluidics and nanofluidics \cite{Vanderpoorten2019}, novel polariton-enhanced chemical sensors can be envisioned.  

There should be also interesting and informative polarization effects for strongly coupled molecules in cavities. Polarization dependent transient absorption is a popular way to measure anisotropy decay and characterize solvation of species in visible absorption and emission studies, as well as Raman and infrared spectroscopy. \cite{Fleming1986,Horng1997,Owrutsky1994,Yi1996}. Longer anisotropy decays result from stronger solvent friction that scales with the solute-solvent interaction strength and can be correlated with other properties such as energy relaxation, energy and charge transfer and isomerization. For an ensemble of strongly coupled molecules inside a cavity, excitations are expected to be delocalized throughout the ensemble due to coherent light-matter coupling, as has been already demonstrated by the observation of long range intracavity energy transfer \cite{Zhong2017,Coles2014}. In some respects, the delocalization of a dipole excitation in a cavity resembles a transition in a degenerate vibrational mode of a free-space molecule because it provides a way to rapidly reorient the transition dipole moment. Anisotropy decay for nondegenerate modes outside a cavity requires the physical rotation of a molecule. Degenerate modes can reorient the direction of an excitation without the molecule having to undergo overall rotation. The initial anisotropy is also more rapidly relaxed, as first shown for W(CO)$_6$ \cite{Tokmakoff1995}. Therefore, strongly coupled molecules in a cavity may exhibit faster anisotropy decay due to rapid dipole reorientation among the coupled ensemble than cavity-free molecular species, and also exhibit a lower initial anisotropy.  Further theory and precise ultrafast cavity measurements would be required to verify this intuition.

Optical resonators are already widely used for enhanced sensing in cavity ring-down spectroscopy \cite{Berden2000}. Here the concentration of a gas-phase absorber is usually  small, and the goal of the technique is to achieve as low a minimum detectable concentration as possible (ppb or lower). Strong coupling studies would represent the opposite extreme in which a high concentration of strong absorbers are loaded into the cavity to identify coherently coupled and quantum optical effects on the spectroscopy, photophysics and molecular reaction dynamics.

Coupling electronic or vibrational bands to surface plasmon polariton resonances (SPRs) is a popular approach to enhanced sensing with surface-enhanced Raman spectroscopy (SERS \cite{Haynes2005}) and surface-enhanced infrared spectroscopy (SEIRA \cite{KUNDU2008}). SERS is a mature field in which many orders of magnitude of enhancement can be achieved for otherwise weak signals. SEIRA is also well established with plasmonic resonances of localized particles or meshes \cite{Jensen2000,Rodriguez2007,Neubrech2012} even though the enhancements tend to be much smaller than in SERS. In a recent, related study, enhanced interactions of molecular vibrationsl with relatively narrow surface phonon polariton resonances (SPhPs) of hexagonal boron nitride (hBN) demonstrated not just enhanced absorption but strong coupling as a result of the similar widths for the vibration and SPhPs \cite{Autore2018}. %This is an example of strong coupling being promoted when the interacting resonances have similar widths. \

Several open questions in the field relate to potential consequences of reaching the strong coupling regime for other degrees of freedom that are not directly involved in light-matter interaction. Consider energy transfer between vibrational polaritons in an infrared cavity and a thermal source or sink that is not coupled to light. The coherent energy exchange between cavity photons and the vibrational motion of molecules in an ensemble may modify the rates of thermal energy flux between the system and a reservoir. Additionally, thermal conductivity under electronic strong coupling has yet to be explored. Novel types of coherent control schemes with fewer requirements and constraints on the systems that can be explored, in comparison with gas phase systems, may also be investigated. The potential for inducing polariton-assisted pathways for enhancing the light-harvesting efficiency of biologically-inspired devices is another interesting possibility \cite{Coles2014-chlorosomes,Vergauwe2016} .  Fundamental questions regarding the effect of strong vibrational coupling in the diffusive transport of dilute molecules in solutions, the macroscopic fluid properties of neat liquids, and the dynamics of non-equilibrium and possibly non-statistical chemical reactions should also be explored.

We would finally like to highlight how in two decades the field of molecular polaritons has reached a stage of maturity where emerging research directions can potentially lead to novel room-temperature devices with  enhanced transport, optical and chemical properties, partly due to the formation of molecular polaritons in the optical or infrared regimes. We believe that in order to materialize this promise, it will be necessary for theoretical and experimental researchers to interact closely and focus on the many open research questions in the field. Addressing these challenges would require the successful integration of concepts and techniques from physical chemistry, quantum optics, quantum control, computational physics and materials science.

\acknowledgements
We thank Blake Simpkins, Adam Dunkelberger, Iv\'{a}n Jara, Federico Hern\'{a}ndez, and Johan Triana for valuable discussions. FH is supported by CONICYT through the Proyecto REDES ETAPA INICIAL, Convocatoria 2017 no. REDI 170423, FONDECYT Regular No. 1181743, and Iniciativa Cient\'{i}fica Milenio (ICM) through the Millennium Institute for Research in Optics (MIRO). JO is supported by the Office of Naval Research through internal funding at
        the U.S. Naval Research Laboratory.

\appendix
\section{Homogeneous quantum model from semiclassical theory}
\label{app:Fmatrix derivation}

Consider an individual molecule with three relevant energy eigenstates denoted $\ket{E_0}$, $\ket{E_1}$, and $\ket{E_2}$. Transitions between these molecular states couple  to a quantized single-mode electromagnetic field with annihilation operator $\hat a$ and frequency $\omega_c$ according to the Hamiltonian (units of $\hbar =1$)
\begin{equation}\label{eq:HQ}
\hat H_Q = \omega_c\hat a^\dagger \hat a +\sum_{n} \omega_n\ket{E_n}\bra{E_n} + \left(\Omega_{01}\ket{E_0}\bra{E_1}+\Omega_{02}\ket{E_0}\bra{E_2}\right)\otimes(\hat a^\dagger+\hat a) +{\rm H. c.},
\end{equation}
where $\omega_m$ are the state energies, and $\Omega_{nm}=\Omega_{mn}$ are the state-dependent Rabi frequencies associated with the cavity-induced transitions $E_{n}\leftrightarrow E_m$ ($n\neq m$). Counter-rotating terms are kept but permanent dipoles moments are ignored. In Hamiltonian $\hat H_Q$ both light and matter degrees of freedom are quantized. Upon truncating the photon number to the vacuum state ($n_c=0$) and the single photon Fock state ($n_c=1$), Eq. (\ref{eq:HQ}) can be written in matrix form as
\begin{equation}
\hat H_Q=\left(\begin{array}{c|ccc|cc}
\omega_0 & 	0		& 	0	& 0	&  \Omega_{01}&  \Omega_{02} \\\hline
 	0	& \omega_1& 		0	& \Omega_{01} &  0& 0 \\
 	0	& 	0		& \omega_2			& \Omega_{02}&0  & 0 \\
	0	 & \Omega_{01}			& 	\Omega_{02}			& \omega_0+\omega_c & 0 & 0 \\\hline
 	 \Omega_{01}	 & 	0			& 		0		& 0 &  \omega_1+\omega_c&  0 \\
	 \Omega_{02}	 & 	0			& 		0		&  0&0  &  \omega_2+\omega_c 
\end{array}\right)
\end{equation}

The upper-right sub-block and its Hermitian conjugate lead to counter-rotating modifications of the ground state $\ket{E_0,n_c=0}$ which can be ignored when $\Omega_{01}\ll (\omega_{10}+\omega_c)$ and $\Omega_{02}\ll (\omega_{20}+\omega_c)$, with transition frequencies $\omega_{mn}\equiv \omega_m-\omega_n$. These conditions impose the rotating-wave approximation (RWA).  The first excited polaritons are thus given by the eigenstates of the center sub-block. Upon subtraction of the ground state energy $\omega_0$ along the diagonal and using $\omega_{20}=\omega_{10}+\delta_{21}$, the rotating-wave sub-block reads
 \begin{equation}\label{eq:HQmat rwa}
\hat H_Q^{\rm RWA}=\left(\begin{array}{ccc}
 	 \omega_{10}& 		0	& \Omega_{01}  \\
 	 	0		& \omega_{10}+\delta_{12}			& \Omega_{02}\\
	 \Omega_{01}			& 	\Omega_{02}			& \omega_c 
\end{array}\right)
\end{equation} 
The two-level version of Eq. (\ref{eq:HQmat rwa}) is obtained by setting $\Omega_{02}=0$. Alternatively, it can be derived  from a truncated  Hopfield model of coupled quantum harmonic oscillators, the Jaynes-Cummings model, or the Tavis-Cummings model in the bosonic (Holstein-Primakoff) approximation. 

Now consider a {\it semiclassical} version of the light-matter Hamiltonian Eq. (\ref{eq:HQ}) in which the single cavity mode at frequency $\omega_c$ is treated as the time-dependent field amplitude $\mathcal{E}_c(t)=\mathcal{E}_0 \cos{\omega_c t}$. We can rewrite the bare molecular frequencies to give the semiclassical Hamiltonian
\begin{eqnarray}
\hat H_{SC}(t) &=& \frac{(\omega_0+\omega_1)}{2}\,\hat 1_{3\times 3} - \frac{\omega_{10}}{2} \ket{E_0}\bra{E_0}+\frac{\omega_{10}}{2} \ket{E_1}\bra{E_1} +\left(\frac{\omega_{10}}{2}+\delta_{21}\right)\ket{E_2}\bra{E_2}\nonumber\\
&&+\left(d_{01} \ket{E_0}\bra{E_1}+ d_{02} \ket{E_0}\bra{E_2} \right)\mathcal{E}_0\cos{\omega_c t} +{\rm H.c.},
\end{eqnarray}
where $d_{nm}=d_{mn}$ are the relevant transition dipole moments ($n\neq m$). The first term is a state-independent matrix that only contributes with a global phase to the system eigenstates and can thus be ignored. The explicit time dependence in $\hat H_{SC}(t)$ can be removed with a unitary (rotating frame)  transformation of the form 
\begin{equation}
\hat U(t) = {\rm e}^{i\lambda_0 t} \ket{E_0}\bra{E_0} + {\rm e}^{i\lambda_1 t}\ket{E_1}\bra{E_1} +{\rm e}^{i\lambda_2 t}\ket{E_2}\bra{E_2},
\end{equation}
with frequency parameters $\lambda_n$ that up to this point can be left undetermined. We define a transformed rotating-frame Hamiltonian in the usual form as $\tilde H_{SC} = \hat U\hat H_{SC}\hat U^\dagger +i\left[\frac{d}{dt}\hat U\right] \hat U^\dagger$, which explicitly reads
\begin{eqnarray}
\tilde H_{SC}(t) &=& \left(- \frac{\omega_{10}}{2} -\lambda_0\right)\ket{E_0}\bra{E_0}+\left(\frac{\omega_{10}}{2} -\lambda_1\right)\ket{E_1}\bra{E_1} +\left(\frac{\omega_{10}}{2}+\delta_{21}-\lambda_2\right)\ket{E_2}\bra{E_2}\nonumber\\
&&+\left(g_{01} \ket{E_0}\bra{E_1}{\rm e}^{i(\lambda_0-\lambda_1)t}+ g_{02} \ket{E_0}\bra{E_2}{\rm e}^{i(\lambda_0-\lambda_2)t} \right)\left({\rm e}^{i\omega_c t}+{\rm e}^{-i\omega_c t} \right)\nonumber\\
&&+\left(g_{01} \ket{E_1}\bra{E_0}{\rm e}^{-i(\lambda_0-\lambda_1)t}+ g_{02} \ket{E_2}\bra{E_0}{\rm e}^{-i(\lambda_0-\lambda_2)t} \right)\left({\rm e}^{i\omega_c t}+{\rm e}^{-i\omega_c t} \right),
\end{eqnarray}
where we have defined Rabi frequencies $g_{nm}\equiv d_{nm}\mathcal{E}_0/2$. We now constrain two of the parameters in $U(t)$ by setting $\lambda_1=\lambda_2$ and $\lambda_1-\lambda_0=\omega_c$, with $\lambda_0$ free, to give
\begin{eqnarray}\label{eq}
\tilde H_{SC}(t) &=& \left(- \frac{\omega_{10}}{2} -\lambda_0\right)\ket{E_0}\bra{E_0}+\left(\frac{\omega_{10}}{2} -\lambda_1\right)\ket{E_1}\bra{E_1} +\left(\frac{\omega_{10}}{2}+\delta_{21}-\lambda_1\right)\ket{E_2}\bra{E_2}\nonumber\\
&&+g_{01} \left(\ket{E_0}\bra{E_1}+\ket{E_1}\bra{E_0}\right)+ g_{02}\left( \ket{E_0}\bra{E_2}+\ket{E_2}\bra{E_0}\right) + \hat V(t)
\end{eqnarray}
where $\hat V(t)$ contains residual terms oscillating at frequencies on the order of $\omega_c+\omega_{10}$ ($|\delta_{21}|\ll \omega_{10}$), which can be ignored in a rotating-wave approximation. We can now fix the remaining phase by setting $\lambda_0=-\omega_c/2$ to arrive at a time-independent Hamiltonian $\tilde H_{SC}^{\rm RWA}\equiv \tilde H_{SC}(t)-\hat V(t)$, given in matrix form by
\begin{equation}\label{eq:Hsc rwa}
\tilde H_{SC}^{\rm RWA}=\frac{1}{2}\left(\begin{array}{ccc}
 	 \omega_{10}-\omega_c& 		0	& 2g_{01}  \\
 	 	0		& \omega_{10}-\omega_c+2\delta_{21}			& 2g_{02}\\
	 2g_{01}			& 	2g_{02}			& -\omega_{10}+\omega_c 
\end{array}\right).
\end{equation} 
This matrix is equivalent to the semiclassical fitting matrix $\mathbf{F}(\omega_c)$ in Eq. (\ref{eq:Fmatrix}) of the main text, after making the identifications: $\ket{E_1}=\ket{E_{\tilde 0}}$, $\ket{E_2}=\ket{E_{\tilde 1}}$, and $\delta_{21}=\omega_{\rm v}$. The equivalence of $\tilde H_{SC}^{\rm RWA}$ with the {\it quantum} Hamiltonian $\hat H_{\rm Q}^{\rm RWA}$ in Eq. (\ref{eq:HQmat rwa}) is established by making the replacement $g_{mn}\rightarrow\Omega_{nm}$ and adding the overall shift $(\omega_{10}+\omega_c)/2$ to the diagonal elements in Eq. (\ref{eq:Hsc rwa}), which again only introduces an irrelevant global phase to the system evolution.  

This simple derivation proves that the quantized nature of a cavity field is {\it irrelevant} to describe the dynamics of the lowest excited polariton eigenstates in the rotating-wave approximation, and therefore most cavity spectroscopy experiments in the linear regime, i.e. a classical field gives the same predictions. This conclusion holds for both the Jaynes-Cummings (single molecule) and the homogeneous Tavis-Cummings model (ensemble), upon truncation to the single-excitation manifold. 

%homogeneous quantum model for molecular polaritons truncated to the single-excitation manifold can be equally described using semiclassical light-matter 

\bibliographystyle{unsrt}
\bibliography{perspective}

\begin{thebibliography}{100}

\bibitem{Scully-book}
Marlan~O. Scully and M.~Suhail Zubairy.
\newblock {\em Quantum Optics}.
\newblock Cambridge University Press, 1997.

\bibitem{Mabuchi2002}
H.~Mabuchi and A.~C. Doherty.
\newblock Cavity quantum electrodynamics: Coherence in context.
\newblock {\em Science}, 298(5597):1372--1377, 2002.

\bibitem{Raimond2001}
J.~M. Raimond, M.~Brune, and S.~Haroche.
\newblock Manipulating quantum entanglement with atoms and photons in a cavity.
\newblock {\em Rev. Mod. Phys.}, 73:565--582, Aug 2001.

\bibitem{Shapiro-Brumer-book}
Moshe Shapiro and Paul Brumer.
\newblock {\em {Quantum Control of Molecular Processes}}.
\newblock Wiley-VCH, 2011.

\bibitem{Tannor1988}
David~J Tannor and Stuart~A Rice.
\newblock Coherent pulse sequence control of product formation in chemical
  reactions.
\newblock {\em Adv. Chem. Phys}, 70:441--523, 1988.

\bibitem{Buluta2011}
Iulia Buluta, Sahel Ashhab, and Franco Nori.
\newblock Natural and artificial atoms for quantum computation.
\newblock {\em Reports on Progress in Physics}, 74(10):104401, sep 2011.

\bibitem{Dantus2004}
Marcos Dantus and Vadim~V. Lozovoy.
\newblock Experimental coherent laser control of physicochemical processes.
\newblock {\em Chemical Reviews}, 104(4):1813--1860, 2004.
\newblock PMID: 15080713.

\bibitem{Sussman2006}
Benjamin~J. Sussman, Dave Townsend, Misha~Yu. Ivanov, and Albert Stolow.
\newblock Dynamic stark control of photochemical processes.
\newblock {\em Science}, 314(5797):278--281, 2006.

\bibitem{Cheng:2009}
Yuan-Chung Cheng and Graham~R Fleming.
\newblock {Dynamics of Light Harvesting in Photosynthesis}.
\newblock {\em Annu. Rev. Phys. Chem.}, 60(1):241--262, 2009.

\bibitem{Weinacht2002}
T~C Weinacht and P~H Bucksbaum.
\newblock Using feedback for coherent control of quantum systems.
\newblock {\em Journal of Optics B: Quantum and Semiclassical Optics},
  4(3):R35--R52, apr 2002.

\bibitem{Brif2010}
Constantin Brif, Raj Chakrabarti, and Herschel Rabitz.
\newblock Control of quantum phenomena: past, present and future.
\newblock {\em New Journal of Physics}, 12(7):075008, jul 2010.

\bibitem{Caruso2012}
F.~Caruso, S.~Montangero, T.~Calarco, S.~F. Huelga, and M.~B. Plenio.
\newblock Coherent optimal control of photosynthetic molecules.
\newblock {\em Phys. Rev. A}, 85:042331, Apr 2012.

\bibitem{Hoyer2014}
Stephan Hoyer, Filippo Caruso, Simone Montangero, Mohan Sarovar, Tommaso
  Calarco, Martin~B Plenio, and K~Birgitta Whaley.
\newblock Realistic and verifiable coherent control of excitonic states in a
  light-harvesting complex.
\newblock {\em New Journal of Physics}, 16(4):045007, apr 2014.

\bibitem{Shore1991}
B.~W. Shore, K.~Bergmann, J.~Oreg, and S.~Rosenwaks.
\newblock Multilevel adiabatic population transfer.
\newblock {\em Phys. Rev. A}, 44:7442--7447, Dec 1991.

\bibitem{Dudovich2005}
Nirit Dudovich, Thomas Polack, Avi Pe'er, and Yaron Silberberg.
\newblock Simple route to strong-field coherent control.
\newblock {\em Phys. Rev. Lett.}, 94:083002, Mar 2005.

\bibitem{Chen2010}
Xi~Chen, A.~Ruschhaupt, S.~Schmidt, A.~del Campo, D.~Gu\'ery-Odelin, and J.~G.
  Muga.
\newblock Fast optimal frictionless atom cooling in harmonic traps: Shortcut to
  adiabaticity.
\newblock {\em Phys. Rev. Lett.}, 104:063002, Feb 2010.

\bibitem{Vitanov2017}
Nikolay~V. Vitanov, Andon~A. Rangelov, Bruce~W. Shore, and Klaas Bergmann.
\newblock Stimulated raman adiabatic passage in physics, chemistry, and beyond.
\newblock {\em Rev. Mod. Phys.}, 89:015006, Mar 2017.

\bibitem{Ospelkaus2008}
S.~Ospelkaus, A.~Pe'er, K.~K. Ni, J.~J. Zirbel, B.~Neyenhuis, S.~Kotochigova,
  P.~S. Julienne, J.~Ye, and D.~S. Jin.
\newblock Efficient state transfer in an ultracold dense gas of heteronuclear
  molecules.
\newblock {\em Nature Physics}, 4(8):622--626, 2008.

\bibitem{Corrales2014}
M.~E. Corrales, J.~Gonz{\'a}lez-V{\'a}zquez, G.~Balerdi, I.~R. Sol{\'a},
  R.~de~Nalda, and L.~Ba{\~n}ares.
\newblock Control of ultrafast molecular photodissociation by
  laser-field-induced potentials.
\newblock {\em Nature Chemistry}, 6:785 EP --, 07 2014.

\bibitem{Gorniaczyk2014}
H.~Gorniaczyk, C.~Tresp, J.~Schmidt, H.~Fedder, and S.~Hofferberth.
\newblock Single-photon transistor mediated by interstate rydberg interactions.
\newblock {\em Phys. Rev. Lett.}, 113:053601, Jul 2014.

\bibitem{Wei2014}
Yu-Jia Wei, Yu-Ming He, Ming-Cheng Chen, Yi-Nan Hu, Yu~He, Dian Wu, Christian
  Schneider, Martin Kamp, Sven H{\"o}fling, Chao-Yang Lu, and Jian-Wei Pan.
\newblock Deterministic and robust generation of single photons from a single
  quantum dot with 99.5\% indistinguishability using adiabatic rapid passage.
\newblock {\em Nano Letters}, 14(11):6515--6519, 2014.
\newblock PMID: 25357153.

\bibitem{Beterov2016}
I.~I. Beterov, M.~Saffman, E.~A. Yakshina, D.~B. Tretyakov, V.~M. Entin,
  S.~Bergamini, E.~A. Kuznetsova, and I.~I. Ryabtsev.
\newblock Two-qubit gates using adiabatic passage of the stark-tuned f\"orster
  resonances in rydberg atoms.
\newblock {\em Phys. Rev. A}, 94:062307, Dec 2016.

\bibitem{Yale2016}
Christopher~G. Yale, F.~Joseph Heremans, Brian~B. Zhou, Adrian Auer, Guido
  Burkard, and David~D. Awschalom.
\newblock Optical manipulation of the berry phase in a solid-state spin qubit.
\newblock {\em Nature Photonics}, 10:184 EP --, 02 2016.

\bibitem{Kumar2016}
K.~S. Kumar, A.~Veps{\"a}l{\"a}inen, S.~Danilin, and G.~S. Paraoanu.
\newblock Stimulated raman adiabatic passage in a three-level superconducting
  circuit.
\newblock {\em Nature Communications}, 7(1):10628, 2016.

\bibitem{Du2016}
Yan-Xiong Du, Zhen-Tao Liang, Yi-Chao Li, Xian-Xian Yue, Qing-Xian Lv, Wei
  Huang, Xi~Chen, Hui Yan, and Shi-Liang Zhu.
\newblock Experimental realization of stimulated raman shortcut-to-adiabatic
  passage with cold atoms.
\newblock {\em Nature Communications}, 7(1):12479, 2016.

\bibitem{Yan2019}
Tongxing Yan, Bao-Jie Liu, Kai Xu, Chao Song, Song Liu, Zhensheng Zhang, Hui
  Deng, Zhiguang Yan, Hao Rong, Keqiang Huang, Man-Hong Yung, Yuanzhen Chen,
  and Dapeng Yu.
\newblock Experimental realization of nonadiabatic shortcut to non-abelian
  geometric gates.
\newblock {\em Phys. Rev. Lett.}, 122:080501, Feb 2019.

\bibitem{Mukamel:2004}
Shaul Mukamel and Darius Abramavicius.
\newblock {Many-Body Approaches for Simulating Coherent Nonlinear
  Spectroscopies of Electronic and Vibrational Excitons}.
\newblock {\em Chem. Rev.}, 104(4):2073--2098, 2004.

\bibitem{Hutchison2012}
James~A Hutchison, Tal Schwartz, Cyriaque Genet, E.~Devaux, and T.~W. Ebbesen.
\newblock {Modifying Chemical Landscapes by Coupling to Vacuum Fields}.
\newblock {\em Angew. Chem. Int. Ed.}, 51(7):1592--1596, 2012.

\bibitem{Schartz2011}
T.~Schwartz, J.~A. Hutchison, C.~Genet, and T.~W. Ebbesen.
\newblock Reversible switching of ultrastrong light-molecule coupling.
\newblock {\em Phys. Rev. Lett.}, 106:196405, 2011.

\bibitem{Orgiu2015}
E.~Orgiu~{\it et al.}
\newblock Conductivity in organic semiconductors hybridized with the vacuum
  field.
\newblock {\em Nat Mater}, 14(11):1123--1129, 11 2015.

\bibitem{Ebbesen2016}
Thomas~W Ebbesen.
\newblock {Hybrid Light-Matter States in a Molecular and Material Science
  Perspective}.
\newblock {\em Accounts of Chemical Research}, 49:2403--2412, 2016.

\bibitem{Barachati2018}
F{\'a}bio Barachati, Janos Simon, Yulia~A. Getmanenko, Stephen Barlow, Seth~R.
  Marder, and St{\'e}phane K{\'e}na-Cohen.
\newblock Tunable third-harmonic generation from polaritons in the ultrastrong
  coupling regime.
\newblock {\em ACS Photonics}, 5(1):119--125, 2018.

\bibitem{Daskalakis2017}
Konstantinos~S Daskalakis, Stefan~A Maier, and St{\'e}phane K{\'e}na-Cohen.
\newblock Polariton condensation in organic semiconductors.
\newblock In {\em Quantum Plasmonics}, pages 151--163. Springer International
  Publishing, 2017.

\bibitem{Kena-Cohen:2008}
S~K{\'e}na-Cohen, M.~{Davan\ifmmode \mbox\c{c}\else \c{c}\fi{}o}, and S.~R.
  Forrest.
\newblock {Strong Exciton-Photon Coupling in an Organic Single Crystal
  Microcavity}.
\newblock {\em Phys. Rev. Lett.}, 101:116401, Sep 2008.

\bibitem{Kena-Cohen:2010}
S.~Kena-Cohen and S.~R. Forrest.
\newblock {Room-Temperature Polariton Lasing in an Organic Single-Crystal
  Microcavity}.
\newblock {\em Nature Photon.}, 4(6):371--375, jun 2010.

\bibitem{Kena-Cohen2013}
St{\'e}phane K{\'e}na-Cohen, Stefan~A. Maier, and Donal D.~C. Bradley.
\newblock Ultrastrongly coupled exciton--polaritons in metal-clad organic
  semiconductor microcavities.
\newblock {\em Advanced Optical Materials}, 1(11):827--833, 2013.

\bibitem{Lerario2017}
Giovanni Lerario, Antonio Fieramosca, F{\'a}bio Barachati, Dario Ballarini,
  Konstantinos~S Daskalakis, Lorenzo Dominici, Milena De~Giorgi, Stefan~A
  Maier, Giuseppe Gigli, and St{\'e}phane K{\'e}na-Cohen.
\newblock Room-temperature superfluidity in a polariton condensate.
\newblock {\em Nature Physics}, 2017.

\bibitem{Long2015}
J.~P. Long and B.~S. Simpkins.
\newblock Coherent coupling between a molecular vibration and fabry--perot
  optical cavity to give hybridized states in the strong coupling limit.
\newblock {\em ACS Photonics}, 2(1):130--136, 2015.

\bibitem{Kapon2017}
Omree Kapon, Rena Yitzhari, Alexander Palatnik, and Yaakov~R. Tischler.
\newblock {Vibrational Strong Light-Matter Coupling Using a Wavelength-Tunable
  Mid-infrared Open Microcavity}.
\newblock {\em Journal of Physical Chemistry C}, 121(34):18845--18853, 2017.

\bibitem{Muallem2016}
Merav Muallem, Alexander Palatnik, Gilbert~D. Nessim, and Yaakov~R. Tischler.
\newblock {Strong Light-Matter Coupling and Hybridization of Molecular
  Vibrations in a Low-Loss Infrared Microcavity}.
\newblock {\em Journal of Physical Chemistry Letters}, 7(11):2002--2008, 2016.

\bibitem{Saurabh2016}
Prasoon Saurabh and Shaul Mukamel.
\newblock Two-dimensional infrared spectroscopy of vibrational polaritons of
  molecules in an optical cavity.
\newblock {\em J. Chem. Phys.}, 144(12), 2016.

\bibitem{Simpkins2015}
B.~S. Simpkins, Kenan~P. Fears, Walter~J. Dressick, Bryan~T. Spann, Adam~D.
  Dunkelberger, and Jeffrey~C. Owrutsky.
\newblock {Spanning Strong to Weak Normal Mode Coupling between Vibrational and
  Fabry--P{\'{e}}rot Cavity Modes through Tuning of Vibrational Absorption
  Strength}.
\newblock {\em ACS Photonics}, 2(10):1460--1467, 2015.

\bibitem{Thomas2016}
Anoop Thomas, Jino George, Atef Shalabney, Marian Dryzhakov, Sreejith~J. Varma,
  Joseph Moran, Thibault Chervy, Xiaolan Zhong, Elo{\"{i}}se Devaux, Cyriaque
  Genet, James~A. Hutchison, and Thomas~W. Ebbesen.
\newblock {Ground-State Chemical Reactivity under Vibrational Coupling to the
  Vacuum Electromagnetic Field}.
\newblock {\em Angewandte Chemie - International Edition}, 55(38):11462--11466,
  2016.

\bibitem{Vergauwe2016}
Robrecht M.~A. Vergauwe, Jino George, Thibault Chervy, James~A. Hutchison, Atef
  Shalabney, Vladimir~Y. Torbeev, and Thomas~W. Ebbesen.
\newblock Quantum strong coupling with protein vibrational modes.
\newblock {\em The Journal of Physical Chemistry Letters}, 7(20):4159--4164,
  2016.
\newblock PMID: 27689759.

\bibitem{Chervy2018}
Thibault Chervy, Anoop Thomas, Elias Akiki, Robrecht M.~A. Vergauwe, Atef
  Shalabney, Jino George, Elo{\"{i}}se Devaux, James~A. Hutchison, Cyriaque
  Genet, and Thomas~W. Ebbesen.
\newblock {Vibro-Polaritonic IR Emission in the Strong Coupling Regime}.
\newblock {\em ACS Photonics}, 5(1):217--224, 2018.

\bibitem{George2015}
Jino George, Atef Shalabney, James~A. Hutchison, Cyriaque Genet, and Thomas~W.
  Ebbesen.
\newblock {Liquid-phase vibrational strong coupling}.
\newblock {\em Journal of Physical Chemistry Letters}, 6(6):1027--1031, 2015.

\bibitem{George2016}
Jino George, Thibault Chervy, Atef Shalabney, Elo{\"{i}}se Devaux, Hidefumi
  Hiura, Cyriaque Genet, and Thomas~W. Ebbesen.
\newblock {Multiple Rabi Splittings under Ultrastrong Vibrational Coupling}.
\newblock {\em Physical Review Letters}, 117(15):153601, 2016.

\bibitem{Hertzog2017}
Manuel Hertzog, Per Rudquist, James~A. Hutchison, Jino George, Thomas~W.
  Ebbesen, and Karl B{\"{o}}rjesson.
\newblock {Voltage-Controlled Switching of Strong Light-Matter Interactions
  using Liquid Crystals}.
\newblock {\em Chemistry - A European Journal}, 23(72):18166--18170, 2017.

\bibitem{Shalabney2015coherent}
A.~Shalabney, J.~George, J.~Hutchison, G.~Pupillo, C.~Genet, and T.~W. Ebbesen.
\newblock {Coherent coupling of molecular resonators with a microcavity mode}.
\newblock {\em Nature Communications}, 6(Umr 7006):1--6, 2015.

\bibitem{Shalabney2015raman}
Atef Shalabney, Jino George, Hidefumi Hiura, James~A. Hutchison, Cyriaque
  Genet, Petra Hellwig, and Thomas~W. Ebbesen.
\newblock {Enhanced Raman Scattering from Vibro-Polariton Hybrid States}.
\newblock {\em Angewandte Chemie International Edition}, 54(27):7971--7975,
  2015.

\bibitem{Dunkelberger2016}
A.~D. Dunkelberger, B.~T. Spann, K.~P. Fears, B.~S. Simpkins, and J.~C.
  Owrutsky.
\newblock {Modified relaxation dynamics and coherent energy exchange in coupled
  vibration-cavity polaritons}.
\newblock {\em Nature Communications}, 7:1--10, 2016.

\bibitem{Xiang2018}
Bo~Xiang, Raphael~F. Ribeiro, Adam~D. Dunkelberger, Jiaxi Wang, Yingmin Li,
  Blake~S. Simpkins, Jeffrey~C. Owrutsky, Joel Yuen-Zhou, and Wei Xiong.
\newblock Two-dimensional infrared spectroscopy of vibrational polaritons.
\newblock {\em Proceedings of the National Academy of Sciences},
  115(19):4845--4850, 2018.

\bibitem{Ahn2018}
Wonmi Ahn, Igor Vurgaftman, Adam~D. Dunkelberger, Jeffrey~C. Owrutsky, and
  Blake~S. Simpkins.
\newblock {Vibrational Strong Coupling Controlled by Spatial Distribution of
  Molecules within the Optical Cavity}.
\newblock {\em ACS Photonics}, 5(1):158--166, 2018.

\bibitem{Dunkelberger2018}
Adam Dunkelberger, Roderick Davidson, Wonmi Ahn, Blake Simpkins, and Jeffrey
  Owrutsky.
\newblock {Ultrafast Transmission Modulation and Recovery via Vibrational
  Strong Coupling}.
\newblock {\em The Journal of Physical Chemistry A}, 122(4):965--971, feb 2018.

\bibitem{Thomas2019}
A.~Thomas, L.~Lethuillier-Karl, K.~Nagarajan, R.~M.~A. Vergauwe, J.~George,
  T.~Chervy, A.~Shalabney, E.~Devaux, C.~Genet, J.~Moran, and T.~W. Ebbesen.
\newblock {Tilting a ground-state reactivity landscape by vibrational strong
  coupling}.
\newblock {\em Science}, 363(6427):615--619, 2019.

\bibitem{Dunkelberger2019}
Adam~D. Dunkelberger, Andrea~B. Grafton, Igor Vurgaftman, {\"O}ney~O. Soykal,
  Thomas~L. Reinecke, Roderick~B. Davidson, Blake~S. Simpkins, and Jeffrey~C.
  Owrutsky.
\newblock Saturable absorption in solution-phase and cavity-coupled tungsten
  hexacarbonyl.
\newblock {\em ACS Photonics}, 0(0):null, 2019.

\bibitem{RMiller2005}
R.~Miller~{\it et al.}
\newblock Trapped atoms in cavity qed: coupling quantized light and matter.
\newblock {\em J. Phys. B: At., Mol. Opt. Phys.}, 38(9):S551, 2005.

\bibitem{Reithmaier2004}
J.~P. Reithmaier, G.~Sek, A.~L\"{o}ffler, C.~Hofmann, S.~Kuhn, S.~Reitzenstein,
  L.~V. Keldysh, V.~D. Kulakovskii, T.~L. Reinecke, and A.~Forchel.
\newblock Strong coupling in a single quantum dot--semiconductor microcavity
  system.
\newblock {\em Nature}, 432(7014):197--200, 2004.

\bibitem{Norris1994}
T.~B. Norris, J.-K. Rhee, C.-Y. Sung, Y.~Arakawa, M.~Nishioka, and C.~Weisbuch.
\newblock Time-resolved vacuum rabi oscillations in a semiconductor quantum
  microcavity.
\newblock {\em Phys. Rev. B}, 50:14663--14666, Nov 1994.

\bibitem{Blais2004}
Alexandre Blais, Ren-Shou Huang, Andreas Wallraff, S.~M. Girvin, and R.~J.
  Schoelkopf.
\newblock Cavity quantum electrodynamics for superconducting electrical
  circuits: An architecture for quantum computation.
\newblock {\em Phys. Rev. A}, 69:062320, Jun 2004.

\bibitem{OBrien2009}
Jeremy~L. O'Brien, Akira Furusawa, and Jelena Vu\v{c}kovi\'{c}.
\newblock Photonic quantum technologies.
\newblock {\em Nature Photonics}, 3(12):687--695, December 2009.

\bibitem{Forn-Diaz2018}
P.~Forn-D{\'{i}}az, L.~Lamata, E.~Rico, J.~Kono, and E.~Solano.
\newblock {Ultrastrong coupling regimes of light-matter interaction}.
\newblock {\em Rev. Mod. Phys.}, 91:25005, 2019.

\bibitem{Kockum2019}
Anton~Frisk Kockum, Adam Miranowicz, Simone {De Liberato}, Salvatore Savasta,
  and Franco Nori.
\newblock {Ultrastrong coupling between light and matter}.
\newblock {\em Nature Reviews Physics}, 1:19--40, 2019.

\bibitem{Tischler2007}
J.~R. Tischler~{\it et al.}
\newblock Solid state cavity qed: Strong coupling in organic thin films.
\newblock {\em Organic Electronics}, 8(2--3):94 -- 113, 2007.

\bibitem{Holmes2007}
R.~J. Holmes and S.~R. Forrest.
\newblock Strong exciton--photon coupling in organic materials.
\newblock {\em Org. Electron.}, 8(2--3):77 -- 93, 2007.

\bibitem{Hertzog2019}
Manuel Hertzog, Mao Wang, J{\"u}rgen Mony, and Karl B{\"o}rjesson.
\newblock Strong light--matter interactions: a new direction within chemistry.
\newblock {\em Chem. Soc. Rev.}, 48:937--961, 2019.

\bibitem{Dovzhenko2018}
D.~S. Dovzhenko, S.~V. Ryabchuk, Yu.~P. Rakovich, and I.~R. Nabiev.
\newblock Light--matter interaction in the strong coupling regime:
  configurations{,} conditions{,} and applications.
\newblock {\em Nanoscale}, 10:3589--3605, 2018.

\bibitem{Torma2015}
P.~T\"{o}rm\"{a} and W.~L. Barnes.
\newblock Strong coupling between surface plasmon polaritons and emitters: a
  review.
\newblock {\em Rep. Prog. Phys.}, 78(1):013901, 2015.

\bibitem{Baranov2017}
Denis~G. Baranov, Martin Wers{\"a}ll, Jorge Cuadra, Tomasz~J. Antosiewicz, and
  Timur Shegai.
\newblock Novel nanostructures and materials for strong light--matter
  interactions.
\newblock {\em ACS Photonics}, 5:24--42, 2018.

\bibitem{Cao2018}
En~Cao, Weihua Lin, Mengtao Sun, Wenjie Liang, and Yuzhi Song.
\newblock {Exciton-plasmon coupling interactions: From principle to
  applications}.
\newblock {\em Nanophotonics}, 7(1):145--167, 2018.

\bibitem{Wang2018}
Hai Wang, Hai-Yu Wang, Hong-Bo Sun, Andrea Cerea, Andrea Toma, Francesco
  De~Angelis, Xin Jin, Luca Razzari, Dan Cojoc, Daniele Catone, Fangcheng
  Huang, and Remo Proietti~Zaccaria.
\newblock Dynamics of strongly coupled hybrid states by transient absorption
  spectroscopy.
\newblock {\em Advanced Functional Materials}, 28(48):1801761, 2018.

\bibitem{Agranovich2005}
V.M. Agranovich and G.C.~La Rocca.
\newblock Electronic excitations in organic microcavities with strong
  light--matter coupling.
\newblock {\em Solid State Communications}, 135(9--10):544 -- 553, 2005.
\newblock Fundamental Optical and Quantum Effects in Condensed Matter.

\bibitem{Litinskaya2006}
M.~Litinskaya, P.~Reineker, and V.M. Agranovich.
\newblock Exciton--polaritons in organic microcavities.
\newblock {\em J. Lumin.}, 119:277 -- 282, 2006.

\bibitem{Michetti2008}
Paolo Michetti and Giuseppe~C. La~Rocca.
\newblock Simulation of {J}-aggregate microcavity photoluminescence.
\newblock {\em Phys. Rev. B}, 77:195301, May 2008.

\bibitem{Herrera2017-review}
Felipe Herrera and Frank~C. Spano.
\newblock Theory of nanoscale organic cavities: The essential role of
  vibration-photon dressed states.
\newblock {\em ACS Photonics}, 5:65--79, 2018.

\bibitem{Ribeiro2018a}
Raphael~F. Ribeiro, Luis~A. Mart{\'{i}}nez-Mart{\'{i}}nez, Matthew Du, Jorge
  Campos-Gonzalez-Angulo, and Joel Yuen-Zhou.
\newblock {Polariton chemistry: controlling molecular dynamics with optical
  cavities}.
\newblock {\em Chemical Science}, 9(30):6325--6339, 2018.

\bibitem{Feist2018}
Johannes Feist, Javier Galego, and Francisco~J. Garcia-Vidal.
\newblock {Polaritonic Chemistry with Organic Molecules}.
\newblock {\em ACS Photonics}, 5(1):205--216, 2018.

\bibitem{Flick2018aa}
Johannes Flick, Nicholas Rivera, and Prineha Narang.
\newblock {Strong light-matter coupling in quantum chemistry and quantum
  photonics}.
\newblock {\em Nanophotonics}, 7(9):1479--1501, 2018.

\bibitem{Ruggenthaler2018}
Michael Ruggenthaler, Nicolas Tancogne-Dejean, Johannes Flick, Heiko Appel, and
  Angel Rubio.
\newblock From a quantum-electrodynamical light--matter description to novel
  spectroscopies.
\newblock {\em Nature Reviews Chemistry}, 2:0118 EP --, 03 2018.

\bibitem{Agranovich-book}
V.~M. Agranovich.
\newblock {\em {Excitations in Organic Solids}}.
\newblock {International Series of Monographs in Physics}. Oxford University
  Press, 2008.

\bibitem{Hopfield1960}
D.~G. Thomas and J.~J. Hopfield.
\newblock {\em Phys. Rev. Lett.}, 5:505, 1961.

\bibitem{Agranovich1962}
V.~M. Agranovich and V.~L. Ginzburg.
\newblock {\em Sov. Phys. Usp.}, 5:323, 1962.

\bibitem{Skolnick1998}
M.~S. Skolnick, T.~A. Fisher, and D.~M. Whittaker.
\newblock Strong coupling phenomena in quantum microcavity structures.
\newblock {\em Semiconductor Science and Technology}, 13(7):645, 1998.

\bibitem{Vahala2003}
K.~J. Vahala.
\newblock Optical microcavities.
\newblock {\em Nature}, 424(6950):839--846, 08 2003.

\bibitem{Agranovich1997}
V.~Agranovich, H.~Benisty, and C.~Weisbuch.
\newblock {Organic and inorganic quantum wells in a microcavity:
  Frenkel-Wannier-Mott excitons hybridization and energy transformation}.
\newblock {\em Solid State Communications}, 102(8):631--636, 1997.

\bibitem{Lidzey1998}
D.~G. Lidzey, D.~D.~C. Bradley, M.~S. Skolnick, T.~Virgili, S.~Walker, and
  D.~M. Whittaker.
\newblock Strong exciton-photon coupling in an organic semiconductor
  microcavity.
\newblock {\em Nature}, 395(6697):53--55, 09 1998.

\bibitem{Lidzey1999}
D.~G. Lidzey~{\it et al.}
\newblock Room temperature polariton emission from strongly coupled organic
  semiconductor microcavities.
\newblock {\em Phys. Rev. Lett.}, 82:3316--3319, 1999.

\bibitem{Hobson2002}
P.~A. Hobson, W.~L. Barnes, D.~G. Lidzey, G.~A. Gehring, D.~M. Whittaker, M.~S.
  Skolnick, and S.~Walker.
\newblock Strong exciton--photon coupling in a low-{Q} all-metal mirror
  microcavity.
\newblock {\em Applied Physics Letters}, 81(19):3519--3521, 2002.

\bibitem{Bellessa2004}
J.~Bellessa, C.~Bonnand, J.~C. Plenet, and J.~Mugnier.
\newblock {Strong Coupling between Surface Plasmons and Excitons in an Organic
  Semiconductor}.
\newblock {\em Phys. Rev. Lett.}, 93:036404, Jul 2004.

\bibitem{Ditinger2005}
J.~Dintinger, S.~Klein, F.~Bustos, W.~L. Barnes, and T.~W. Ebbesen.
\newblock {Strong Coupling Between Surface Plasmon-Polaritons and Organic
  Molecules in Subwavelength Hole Arrays}.
\newblock {\em Phys. Rev. B}, 71:035424, Jan 2005.

\bibitem{Gonzalez2013}
A.~Gonz\'alez-Tudela~{\it et al.}
\newblock Theory of strong coupling between quantum emitters and propagating
  surface plasmons.
\newblock {\em Phys. Rev. Lett.}, 110:126801, Mar 2013.

\bibitem{Cacciola2014}
A.~Cacciola~{\it et al.}
\newblock Ultrastrong coupling of plasmons and excitons in a nanoshell.
\newblock {\em ACS Nano}, 8(11):11483--11492, 2014.

\bibitem{Bellessa2014}
J.~Bellessa~{\it et al.}
\newblock Strong coupling between plasmons and organic semiconductors.
\newblock {\em Electronics}, 3(2):303, 2014.

\bibitem{Delga2014}
A~Delga, J~Feist, J~Bravo-Abad, and F~J Garcia-Vidal.
\newblock Theory of strong coupling between quantum emitters and localized
  surface plasmons.
\newblock {\em Journal of Optics}, 16(11):114018, 2014.

\bibitem{Chikkaraddy2016}
Rohit Chikkaraddy, Bart de~Nijs, Felix Benz, Steven~J. Barrow, Oren~A.
  Scherman, Edina Rosta, Angela Demetriadou, Peter Fox, Ortwin Hess, and
  Jeremy~J. Baumberg.
\newblock Single-molecule strong coupling at room temperature in plasmonic
  nanocavities.
\newblock {\em Nature}, 535(7610):127--130, 07 2016.

\bibitem{Benz2016}
Felix Benz, Mikolaj~K. Schmidt, Alexander Dreismann, Rohit Chikkaraddy, Yao
  Zhang, Angela Demetriadou, Cloudy Carnegie, Hamid Ohadi, Bart de~Nijs, Ruben
  Esteban, Javier Aizpurua, and Jeremy~J. Baumberg.
\newblock Single-molecule optomechanics in picocavities.
\newblock {\em Science}, 354(6313):726--729, 2016.

\bibitem{Chikkaraddy2018}
Rohit Chikkaraddy, V.~A. Turek, Nuttawut Kongsuwan, Felix Benz, Cloudy
  Carnegie, Tim van~de Goor, Bart de~Nijs, Angela Demetriadou, Ortwin Hess,
  Ulrich~F. Keyser, and Jeremy~J. Baumberg.
\newblock Mapping nanoscale hotspots with single-molecule emitters assembled
  into plasmonic nanocavities using dna origami.
\newblock {\em Nano Letters}, 18(1):405--411, 2018.
\newblock PMID: 29166033.

\bibitem{Wang2017}
Daqing Wang, Hrishikesh Kelkar, Diego Martin-Cano, Tobias Utikal, Stephan
  G{\"{o}}tzinger, and Vahid Sandoghdar.
\newblock {Coherent Coupling of a Single Molecule to a Scanning Fabry-Perot
  Microcavity}.
\newblock {\em Physical Review X}, 7(2):021014, apr 2017.

\bibitem{Tame2013}
M.~S. Tame, K.~R. McEnery, {\c S}.~K. {\"{O}}zdemir, J.~Lee, S.~a. Maier, and
  M.~S. Kim.
\newblock {Quantum plasmonics}.
\newblock {\em Nature Physics}, 9(6):329--340, jun 2013.

\bibitem{Vanderpoorten2019}
Oliver Vanderpoorten, Quentin Peter, Pavan~K. Challa, Ulrich~F. Keyser, Jeremy
  Baumberg, Clemens~F. Kaminski, and Tuomas P.~J. Knowles.
\newblock Scalable integration of nano-, and microfluidics with hybrid
  two-photon lithography.
\newblock {\em Microsystems \& Nanoengineering}, 5(1):40, 2019.

\bibitem{Li2017}
Yuzhang Li, Yanbin Li, Allen Pei, Kai Yan, Yongming Sun, Chun-Lan Wu,
  Lydia-Marie Joubert, Richard Chin, Ai~Leen Koh, Yi~Yu, John Perrino, Benjamin
  Butz, Steven Chu, and Yi~Cui.
\newblock Atomic structure of sensitive battery materials and interfaces
  revealed by cryo{\textendash}electron microscopy.
\newblock {\em Science}, 358(6362):506--510, 2017.

\bibitem{Dorfman2016}
Konstantin~E. Dorfman, Frank Schlawin, and Shaul Mukamel.
\newblock Nonlinear optical signals and spectroscopy with quantum light.
\newblock {\em Rev. Mod. Phys.}, 88:045008, Dec 2016.

\bibitem{Raymer2013}
M.~G. Raymer, Andrew~H. Marcus, Julia~R. Widom, and Dashiell L.~P. Vitullo.
\newblock Entangled photon-pair two-dimensional fluorescence spectroscopy
  (epp-2dfs).
\newblock {\em The Journal of Physical Chemistry B}, 117(49):15559--15575,
  2013.
\newblock PMID: 24047447.

\bibitem{Eizner2019}
Elad Eizner, Luis~A. Mart{\'\i}nez-Mart{\'\i}nez, Joel Yuen-Zhou, and
  St{\'e}phane K{\'e}na-Cohen.
\newblock Inverting singlet and triplet excited states using strong
  light-matter coupling, 2019.

\bibitem{Buhmann2007}
Stefan~Yoshi Buhmann and Dirk-Gunnar Welsch.
\newblock Dispersion forces in macroscopic quantum electrodynamics.
\newblock {\em Progress in Quantum Electronics}, 31(2):51 -- 130, 2007.

\bibitem{Macleod2013}
H.~Angus Macleod.
\newblock {\em {Thin-Film Optical Filters}}.
\newblock CRC Press, 4th edition, 2010.

\bibitem{Fleischhauer:2005}
M.~Fleischhauer, A.~Imamoglu, and J.~P. Marangos.
\newblock {Electromagnetically Induced Transparency: Optics in Coherent Media}.
\newblock {\em Rev. Mod. Phys.}, 77:633--673, Jul 2005.

\bibitem{JohnsonChristy}
P.~B. Johnson and R.~W. Christy.
\newblock Optical constants of the noble metals.
\newblock {\em Phys. Rev. B}, 6:4370--4379, Dec 1972.

\bibitem{Zengin:2013}
G.~Zengin, G.~Johansson, P.~Johansson, T.~J. Antosiewicz, M.~K{\"a}ll, and
  T.~Shegai.
\newblock {Approaching the Strong Coupling Limit in Single Plasmonic Nanorods
  Interacting with J-Aggregates}.
\newblock {\em Sci. Rep.}, 3:3074, oct 2013.

\bibitem{Boyd-book}
Robert~W Boyd.
\newblock {\em {Nonlinear Optics}}.
\newblock Elsevier, Burlington, MA USA, 3rd edition, 2008.

\bibitem{PALIK1997book}
Edward~D. Palik.
\newblock {\em Handbook of Optical Constants of Solids}.
\newblock Academic Press, Burlington, 1997.

\bibitem{Taflove2005}
Allen Taflove and Susan~C Hagness.
\newblock {\em Computational Electrodynamics}.
\newblock Third edition edition, 2005.

\bibitem{Chantry1982}
G~W Chantry.
\newblock The use of fabry-perot interferometers, etalons and resonators at
  infrared and longer wavelengths-an overview.
\newblock {\em Journal of Physics E: Scientific Instruments}, 15(1):3--8, jan
  1982.

\bibitem{Li2016-plasmons}
Rui-Qi Li, D.~Hern\'angomez-P\'erez, F.~J. Garc\'{\i}a-Vidal, and A.~I.
  Fern\'andez-Dom\'{\i}nguez.
\newblock Transformation optics approach to plasmon-exciton strong coupling in
  nanocavities.
\newblock {\em Phys. Rev. Lett.}, 117:107401, Aug 2016.

\bibitem{Li2018}
Rui-Qi Li, F.~J. Garc{\'\i}a-Vidal, and A.~I. Fern{\'a}ndez-Dom{\'\i}nguez.
\newblock Plasmon-exciton coupling in symmetry-broken nanocavities.
\newblock {\em ACS Photonics}, 5(1):177--185, 2018.

\bibitem{Barnett-Radmore}
Stephen~M. Barnett and Paul Radmore.
\newblock {\em Methods in theoretical quantum optics}.
\newblock Oxford University Press, 1997.

\bibitem{Spano2010}
Frank~C Spano.
\newblock The spectral signatures of frenkel polarons in h- and j-aggregates.
\newblock {\em Acc. Chem. Res.}, 43(3):429--439, 2010.

\bibitem{Herrera2017-PRA}
Felipe Herrera and Frank~C. Spano.
\newblock Absorption and photoluminescence in organic cavity qed.
\newblock {\em Phys. Rev. A}, 95:053867, May 2017.

\bibitem{Carmichael-book2}
H~Carmichael.
\newblock {\em Statistical Mehods in Quantum Optics 2: Non-Classical Fields}.
\newblock Springer Berlin / Heidelberg, 2008.

\bibitem{Hernandez2019}
Federico~J. Hern{\'a}ndez and Felipe Herrera.
\newblock Multi-level quantum rabi model for anharmonic vibrational polaritons.
\newblock {\em The Journal of Chemical Physics}, 151(14):144116, 2019.

\bibitem{Andrews2018}
David~L. Andrews, Garth~A. Jones, A.~Salam, and R.~Guy Woolley.
\newblock Perspective: Quantum hamiltonians for optical interactions.
\newblock {\em The Journal of Chemical Physics}, 148(4):040901, 2018.

\bibitem{Litinskaya2006-disorder}
Marina Litinskaya and Peter Reineker.
\newblock Loss of coherence of exciton polaritons in inhomogeneous organic
  microcavities.
\newblock {\em Phys. Rev. B}, 74:165320, Oct 2006.

\bibitem{Spano1989a}
Frank~C. Spano and Shaul Mukamel.
\newblock {Superradiance in molecular aggregates}.
\newblock {\em The Journal of Chemical Physics}, 91(2):683, 1989.

\bibitem{TANJISUZUKI2011}
Haruka Tanji-Suzuki, Ian~D. Leroux, Monika~H. Schleier-Smith, Marko Cetina,
  Andrew~T. Grier, Jonathan Simon, and Vladan Vuleti{\'c}.
\newblock Chapter 4 - interaction between atomic ensembles and optical
  resonators: Classical description.
\newblock In E.~Arimondo, P.R. Berman, and C.C. Lin, editors, {\em Advances in
  Atomic, Molecular, and Optical Physics}, volume~60 of {\em Advances In
  Atomic, Molecular, and Optical Physics}, pages 201 -- 237. Academic Press,
  2011.

\bibitem{SAVONA1995}
V.~Savona, L.C. Andreani, P.~Schwendimann, and A.~Quattropani.
\newblock Quantum well excitons in semiconductor microcavities: Unified
  treatment of weak and strong coupling regimes.
\newblock {\em Solid State Communications}, 93(9):733 -- 739, 1995.

\bibitem{Stuart2007}
Christina~M. Stuart, Renee~R. Frontiera, and Richard~A. Mathies.
\newblock {Excited-state structure and dynamics of cis- and trans-Azobenzene
  from resonance Raman intensity analysis}.
\newblock {\em Journal of Physical Chemistry A}, 111(48):12072--12080, 2007.

\bibitem{Litinskaya2019}
Marina Litinskaya and Felipe Herrera.
\newblock Vacuum-enhanced optical nonlinearities with disordered molecular
  photoswitches.
\newblock {\em Phys. Rev. B}, 99:041107, Jan 2019.

\bibitem{May-Kuhn}
Volkhard May and Oliver Kuhn.
\newblock {\em {Charge and Energy Transfer Dynamics in Molecular Systems}},
  volume~35.
\newblock Wiley-VCH Verlag GmbH {\&} Co., 3rd edition, 2011.

\bibitem{Wurthner2011}
F.~W{\"u}rthner, Th.~E. Kaise, and Ch.~R. Saha-M{\"o}ller.
\newblock {J-Aggregates: From Serendipitous Discovery to Supramolecular
  Engineering of Functional Dye Materials.}
\newblock {\em Angew. Chem. Int. Ed.}, 50:3376--410, 2011.

\bibitem{Hestand2018}
Nicholas~J. Hestand and Frank~C. Spano.
\newblock Expanded theory of h- and j-molecular aggregates: The effects of
  vibronic coupling and intermolecular charge transfer.
\newblock {\em Chemical Reviews}, 118(15):7069--7163, 2018.
\newblock PMID: 29664617.

\bibitem{Litinskaya2004}
M.~Litinskaya, P.~Reineker, and V.~M. Agranovich.
\newblock Fast polariton relaxation in strongly coupled organic microcavities.
\newblock {\em Journal of Luminescence}, 110(4 SPEC. ISS.):364--372, 2004.

\bibitem{Lopez2007}
C.~E. L\'opez, H.~Christ, J.~C. Retamal, and E.~Solano.
\newblock Effective quantum dynamics of interacting systems with inhomogeneous
  coupling.
\newblock {\em Phys. Rev. A}, 75:033818, Mar 2007.

\bibitem{Spano2015}
F.~C. Spano.
\newblock Optical microcavities enhance the exciton coherence length and
  eliminate vibronic coupling in j-aggregates.
\newblock {\em J. Chem. Phys.}, 142(18):184707, 2015.

\bibitem{Herrera2016}
Felipe Herrera and Frank~C. Spano.
\newblock Cavity-controlled chemistry in molecular ensembles.
\newblock {\em Phys. Rev. Lett.}, 116:238301, Jun 2016.

\bibitem{Shammah2017}
Nathan Shammah, Neill Lambert, Franco Nori, and Simone De~Liberato.
\newblock Superradiance with local phase-breaking effects.
\newblock {\em Phys. Rev. A}, 96:023863, Aug 2017.

\bibitem{Herrera2017-PRL}
Felipe Herrera and Frank~C. Spano.
\newblock Dark vibronic polaritons and the spectroscopy of organic
  microcavities.
\newblock {\em Phys. Rev. Lett.}, 118:223601, May 2017.

\bibitem{Zhong2016}
Xiaolan Zhong, Thibault Chervy, Shaojun Wang, Jino George, Anoop Thomas,
  James~A. Hutchison, Eloise Devaux, Cyriaque Genet, and Thomas~W. Ebbesen.
\newblock Non-radiative energy transfer mediated by hybrid light-matter states.
\newblock {\em Angewandte Chemie International Edition}, 55(21):6202--6206,
  2016.

\bibitem{Zhong2017}
Xiaolan Zhong, Thibault Chervy, Lei Zhang, Anoop Thomas, Jino George, Cyriaque
  Genet, James~A. Hutchison, and Thomas~W. Ebbesen.
\newblock Energy transfer between spatially separated entangled molecules.
\newblock {\em Angewandte Chemie International Edition}, 56(31):9034--9038,
  2017.

\bibitem{Georgiou2018}
Kyriacos Georgiou, Paolo Michetti, Lizhi Gai, Marco Cavazzini, Zhen Shen, and
  David~G. Lidzey.
\newblock Control over energy transfer between fluorescent bodipy dyes in a
  strongly coupled microcavity.
\newblock {\em ACS Photonics}, 5(1):258--266, 2018.

\bibitem{Coles2014-chlorosomes}
David~M. Coles, Yanshen Yang, Yaya Wang, Richard~T. Grant, Robert~A. Taylor,
  Semion~K. Saikin, Al{\'a}n Aspuru-Guzik, David~G. Lidzey, Joseph Kuo-Hsiang
  Tang, and Jason~M. Smith.
\newblock Strong coupling between chlorosomes of photosynthetic bacteria and a
  confined optical cavity mode.
\newblock {\em Nature Communications}, 5(1):5561, 2014.

\bibitem{Grant2018}
Richard~T. Grant, Rahul Jayaprakash, David~M Coles, Andrew Musser, Simone {De
  Liberato}, Ifor~D.W. Samuel, Graham~A. Turnbull, Jenny Clark, and David~G.
  Lidzey.
\newblock {Strong coupling in a microcavity containing $\beta$-carotene}.
\newblock {\em Optics Express}, 26(3):3320, 2018.

\bibitem{Vergauwe2019}
Robrecht M.~A. Vergauwe, Anoop Thomas, Kalaivanan Nagarajan, Atef Shalabney,
  Jino George, Thibault Chervy, Marcus Seidel, Elo{\"\i}se Devaux, Vladimir
  Torbeev, and Thomas~W. Ebbesen.
\newblock Modification of enzyme activity by vibrational strong coupling of
  water.
\newblock {\em Angewandte Chemie International Edition}, 58(43):15324--15328,
  2019.

\bibitem{Takahashi2019}
Shota Takahashi, Kazuya Watanabe, and Yoshiyasu Matsumoto.
\newblock Singlet fission of amorphous rubrene modulated by polariton
  formation.
\newblock {\em The Journal of Chemical Physics}, 151(7):074703, 2019.

\bibitem{Canaguier-Durand2013}
Antoine Canaguier-Durand, Elo{\"{i}}se Devaux, Jino George, Yantao Pang,
  James~A. Hutchison, Tal Schwartz, Cyriaque Genet, Nadine Wilhelms, Jean~Marie
  Lehn, and Thomas~W. Ebbesen.
\newblock {Thermodynamics of molecules strongly coupled to the vacuum field}.
\newblock {\em Angewandte Chemie - International Edition}, 52(40):10533--10536,
  2013.

\bibitem{Stranius2018}
Kati Stranius, Manuel Hertzog, and Karl B{\"o}rjesson.
\newblock Selective manipulation of electronically excited states through
  strong light--matter interactions.
\newblock {\em Nature Communications}, 9(1):2273, 2018.

\bibitem{Ojambati2019}
Oluwafemi~S. Ojambati, Rohit Chikkaraddy, William~D. Deacon, Matthew Horton,
  Dean Kos, Vladimir~A. Turek, Ulrich~F. Keyser, and Jeremy~J. Baumberg.
\newblock Quantum electrodynamics at room temperature coupling a single
  vibrating molecule with a plasmonic nanocavity.
\newblock {\em Nature Communications}, 10(1):1049, 2019.

\bibitem{Held2018}
Martin Held, Arko Graf, Yuriy Zakharko, Pengning Chao, Laura Tropf, Malte~C.
  Gather, and Jana Zaumseil.
\newblock Ultrastrong coupling of electrically pumped near-infrared
  exciton-polaritons in high mobility polymers.
\newblock {\em Advanced Optical Materials}, 6(3):1700962, 2018.

\bibitem{Jayaprakash2019}
Rahul Jayaprakash, Kyriacos Georgiou, Harriet Coulthard, Alexis Askitopoulos,
  Sai~K. Rajendran, David~M. Coles, Andrew~J. Musser, Jenny Clark, Ifor D.~W.
  Samuel, Graham~A. Turnbull, Pavlos~G. Lagoudakis, and David~G. Lidzey.
\newblock A hybrid organic--inorganic polariton led.
\newblock {\em Light: Science \& Applications}, 8(1):81, 2019.

\bibitem{Mazzeo2014}
M.~Mazzeo~{\it et al.}
\newblock Ultrastrong light-matter coupling in electrically doped microcavity
  organic light emitting diodes.
\newblock {\em Appl. Phys. Lett.}, 104(23), 2014.

\bibitem{Gambino2015}
S.~Gambino~{\it et al.}
\newblock Ultrastrong light-matter coupling in electroluminescent organic
  microcavities.
\newblock {\em Applied Materials Today}, 1(1):33 -- 36, 2015.

\bibitem{Balci2013}
S.~Balci.
\newblock Ultrastrong plasmon-exciton coupling in metal nanoprisms with
  j-aggregates.
\newblock {\em Opt. Lett.}, 38(21):4498, 2013.

\bibitem{Erwin2019}
Justin~D. Erwin, Madeline Smotzer, and James~V. Coe.
\newblock Effect of strongly coupled vibration--cavity polaritons on the bulk
  vibrational states within a wavelength-scale cavity.
\newblock {\em The Journal of Physical Chemistry B}, 123(6):1302--1306, 02
  2019.

\bibitem{Lather2019}
Jyoti Lather, Pooja Bhatt, Anoop Thomas, Thomas~W. Ebbesen, and Jino George.
\newblock Cavity catalysis by cooperative vibrational strong coupling of
  reactant and solvent molecules.
\newblock {\em Angewandte Chemie International Edition}, 58(31):10635--10638,
  2019.

\bibitem{Galego2016}
Javier Galego, Francisco~J Garcia-Vidal, and Johannes Feist.
\newblock Suppressing photochemical reactions with quantized light fields.
\newblock {\em Nature Communications}, 7:13841, 2016.

\bibitem{Galego2017}
Javier Galego, Francisco~J. Garcia-Vidal, and Johannes Feist.
\newblock {Many-Molecule Reaction Triggered by a Single Photon in Polaritonic
  Chemistry}.
\newblock {\em Physical Review Letters}, 119(13):1--6, 2017.

\bibitem{Martinez-Martinez2018}
Luis~A. Mart{\'{i}}nez-Mart{\'{i}}nez, Matthew Du, Raphael~F. Ribeiro,
  St{\'{e}}phane K{\'{e}}na-Cohen, and Joel Yuen-Zhou.
\newblock {Polariton-Assisted Singlet Fission in Acene Aggregates}.
\newblock {\em Journal of Physical Chemistry Letters}, 9(8):1951--1957, 2018.

\bibitem{Semenov2019}
Alexander Semenov and Abraham Nitzan.
\newblock Electron transfer in confined electromagnetic fields.
\newblock {\em The Journal of Chemical Physics}, 150(17):174122, 2019.

\bibitem{Du2018-PARET}
Matthew Du, Luis~A. Mart{\'\i}nez-Mart{\'\i}nez, Raphael~F. Ribeiro, Zixuan Hu,
  Vinod~M. Menon, and Joel Yuen-Zhou.
\newblock Theory for polariton-assisted remote energy transfer.
\newblock {\em Chem. Sci.}, 9:6659--6669, 2018.

\bibitem{Du2019}
Matthew Du, Raphael~F. Ribeiro, and Joel Yuen-Zhou.
\newblock Remote control of chemistry in optical cavities.
\newblock {\em Chem}, 5(5):1167 -- 1181, 2019.

\bibitem{Feist2015}
J.~Feist and F.~J. Garcia-Vidal.
\newblock Extraordinary exciton conductance induced by strong coupling.
\newblock {\em Phys. Rev. Lett.}, 114:196402, 2015.

\bibitem{Schachenmayer2015}
J.~Schachenmayer, C.~Genes, E.~Tignone, and G.~Pupillo.
\newblock Cavity-enhanced transport of excitons.
\newblock {\em Phys. Rev. Lett.}, 114:196403, 2015.

\bibitem{Hagenmuller2017}
David Hagenm\"uller, Johannes Schachenmayer, Stefan Sch\"utz, Claudiu Genes,
  and Guido Pupillo.
\newblock Cavity-enhanced transport of charge.
\newblock {\em Phys. Rev. Lett.}, 119:223601, Nov 2017.

\bibitem{Hegenmuller2018}
David Hagenm\"uller, Stefan Sch\"utz, Johannes Schachenmayer, Claudiu Genes,
  and Guido Pupillo.
\newblock Cavity-assisted mesoscopic transport of fermions: Coherent and
  dissipative dynamics.
\newblock {\em Phys. Rev. B}, 97:205303, May 2018.

\bibitem{Mazza2009}
L.~Mazza, L.~Fontanesi, and G.~C. La~Rocca.
\newblock Organic-based microcavities with vibronic progressions:
  Photoluminescence.
\newblock {\em Phys. Rev. B}, 80:235314, Dec 2009.

\bibitem{Cwik2016}
Justyna~A. {\'{C}}wik, Peter Kirton, Simone {De Liberato}, and Jonathan
  Keeling.
\newblock {Excitonic spectral features in strongly coupled organic polaritons}.
\newblock {\em Physical Review A}, 93(3):033840, 2016.

\bibitem{DelPino2015}
Javier del Pino, Johannes Feist, and Francisco~J. Garcia-Vidal.
\newblock {Quantum theory of collective strong coupling of molecular vibrations
  with a microcavity mode}.
\newblock 2015.

\bibitem{delPino2015raman}
Javier del Pino, Johannes Feist, and F.~J. Garcia-Vidal.
\newblock Signatures of vibrational strong coupling in raman scattering.
\newblock {\em The Journal of Physical Chemistry C}, 119(52):29132--29137, 12
  2015.

\bibitem{Strashko2016}
Artem Strashko and Jonathan Keeling.
\newblock Raman scattering with strongly coupled vibron-polaritons.
\newblock {\em arXiv:1606.08343}, 2016.

\bibitem{Ribeiro2018}
Raphael F.~Ribeiro, Adam~D. Dunkelberger, Bo~Xiang, Wei Xiong, Blake~S.
  Simpkins, Jeffrey~C. Owrutsky, and Joel Yuen-Zhou.
\newblock Theory for nonlinear spectroscopy of vibrational polaritons.
\newblock {\em The Journal of Physical Chemistry Letters}, 9(13):3766--3771, 07
  2018.

\bibitem{Cwik2014}
Justyna~A. {\'C}wik, Sahinur Reja, Peter~B. Littlewood, and Jonathan Keeling.
\newblock Polariton condensation with saturable molecules dressed by
  vibrational modes.
\newblock {\em EPL (Europhysics Letters)}, 105(4):47009, 2014.

\bibitem{Strashko2018}
Artem Strashko, Peter Kirton, and Jonathan Keeling.
\newblock Organic polariton lasing and the weak to strong coupling crossover.
\newblock {\em Phys. Rev. Lett.}, 121:193601, Nov 2018.

\bibitem{Herrera2014}
F.~Herrera~{\it et al.}
\newblock Quantum nonlinear optics with polar {J}-aggregates in microcavities.
\newblock {\em J. Phys. Chem. Lett.}, 5(21):3708--3715, 2014.

\bibitem{Flick2017}
Johannes Flick, Michael Ruggenthaler, Heiko Appel, and Angel Rubio.
\newblock Atoms and molecules in cavities, from weak to strong coupling in
  quantum-electrodynamics (qed) chemistry.
\newblock {\em Proceedings of the National Academy of Sciences},
  114(12):3026--3034, 2017.

\bibitem{Vendrell2018}
Oriol Vendrell.
\newblock Coherent dynamics in cavity femtochemistry: Application of the
  multi-configuration time-dependent hartree method.
\newblock {\em Chemical Physics}, 509:55 -- 65, 2018.
\newblock High-dimensional quantum dynamics (on the occasion of the 70th
  birthday of Hans-Dieter Meyer).

\bibitem{Galego2015}
J.~Galego, F.~J. Garcia-Vidal, and J.~Feist.
\newblock Cavity-induced modifications of molecular structure in the
  strong-coupling regime.
\newblock {\em Phys. Rev. X}, 5:041022, Nov 2015.

\bibitem{Luk2017}
Hoi~Ling Luk, Johannes Feist, J.~Jussi Toppari, and Gerrit Groenhof.
\newblock {Multiscale Molecular Dynamics Simulations of Polaritonic Chemistry}.
\newblock {\em Journal of Chemical Theory and Computation}, 13(9):4324--4335,
  2017.

\bibitem{Flick2017a}
Johannes Flick, Heiko Appel, Michael Ruggenthaler, and Angel Rubio.
\newblock {Cavity Born-Oppenheimer Approximation for Correlated
  Electron-Nuclear-Photon Systems}.
\newblock {\em Journal of Chemical Theory and Computation}, 13(4):1616--1625,
  2017.

\bibitem{Flick2018}
Johannes Flick and Prineha Narang.
\newblock Cavity-correlated electron-nuclear dynamics from first principles.
\newblock {\em Phys. Rev. Lett.}, 121:113002, Sep 2018.

\bibitem{Flick2018a}
Johannes Flick, Christian Sch{\"{a}}fer, Michael Ruggenthaler, Heiko Appel, and
  Angel Rubio.
\newblock {Ab Initio Optimized Effective Potentials for Real Molecules in
  Optical Cavities: Photon Contributions to the Molecular Ground State}.
\newblock {\em ACS Photonics}, 5(3):992--1005, 2018.

\bibitem{Vendrell2018aa}
Oriol Vendrell.
\newblock Collective jahn-teller interactions through light-matter coupling in
  a cavity.
\newblock {\em Phys. Rev. Lett.}, 121:253001, Dec 2018.

\bibitem{Fregoni2018}
J.~Fregoni, G.~Granucci, E.~Coccia, M.~Persico, and S.~Corni.
\newblock Manipulating azobenzene photoisomerization through strong
  light--molecule coupling.
\newblock {\em Nature Communications}, 9(1):4688, 2018.

\bibitem{Galego2019}
Javier Galego, Cl\`audia Climent, Francisco~J. Garcia-Vidal, and Johannes
  Feist.
\newblock Cavity casimir-polder forces and their effects in ground-state
  chemical reactivity.
\newblock {\em Phys. Rev. X}, 9:021057, Jun 2019.

\bibitem{Rivera2019}
Nicholas Rivera, Johannes Flick, and Prineha Narang.
\newblock Variational theory of nonrelativistic quantum electrodynamics.
\newblock {\em Phys. Rev. Lett.}, 122:193603, May 2019.

\bibitem{Triana2018}
Johan~F. Triana, Daniel Pel{\'{a}}ez, and Jos{\'{e}}~Luis Sanz-Vicario.
\newblock {Entangled Photonic-Nuclear Molecular Dynamics of LiF in Quantum
  Optical Cavities}.
\newblock {\em Journal of Physical Chemistry A}, 122(8):2266--2278, 2018.

\bibitem{Triana2019}
Johan~F. Triana and Jos\'e~Luis Sanz-Vicario.
\newblock Revealing the presence of potential crossings in diatomics induced by
  quantum cavity radiation.
\newblock {\em Phys. Rev. Lett.}, 122:063603, Feb 2019.

\bibitem{Rossi2019}
Tuomas~P. Rossi, Timur Shegai, Paul Erhart, and Tomasz~J. Antosiewicz.
\newblock Strong plasmon-molecule coupling at the nanoscale revealed by
  first-principles modeling.
\newblock {\em Nature Communications}, 10(1):3336, 2019.

\bibitem{delPino2018}
Javier del Pino, Florian A. Y.~N. Schr\"oder, Alex~W. Chin, Johannes Feist, and
  Francisco~J. Garcia-Vidal.
\newblock Tensor network simulation of non-markovian dynamics in organic
  polaritons.
\newblock {\em Phys. Rev. Lett.}, 121:227401, Nov 2018.

\bibitem{Varguet2019}
H~Varguet, B~Rousseaux, D~Dzsotjan, H~R Jauslin, S~Gu{\'{e}}rin, and G~Colas
  des Francs.
\newblock Non-hermitian hamiltonian description for quantum plasmonics: from
  dissipative dressed atom picture to fano states.
\newblock {\em Journal of Physics B: Atomic, Molecular and Optical Physics},
  52(5):055404, feb 2019.

\bibitem{Reitz2019}
Michael Reitz, Christian Sommer, and Claudiu Genes.
\newblock Langevin approach to quantum optics with molecules.
\newblock {\em Phys. Rev. Lett.}, 122:203602, May 2019.

\bibitem{Imran2019}
Iffat Imran, Giulia~E. Nicolai, Nicholas~D. Stavinski, and Justin~R. Sparks.
\newblock Tuning vibrational strong coupling with co-resonators.
\newblock {\em ACS Photonics}, 6(10):2405--2412, 2019.

\bibitem{Xiang2019manipulating}
Bo~Xiang, Raphael~F. Ribeiro, Yingmin Li, Adam~D. Dunkelberger, Blake~B.
  Simpkins, Joel Yuen-Zhou, and Wei Xiong.
\newblock Manipulating optical nonlinearities of molecular polaritons by
  delocalization.
\newblock {\em Science Advances}, 5(9), 2019.

\bibitem{Xiang2019state}
Bo~Xiang, Raphael~F. Ribeiro, Liying Chen, Jiaxi Wang, Matthew Du, Joel
  Yuen-Zhou, and Wei Xiong.
\newblock State-selective polariton to dark state relaxation dynamics.
\newblock {\em The Journal of Physical Chemistry A}, 123(28):5918--5927, 2019.
\newblock PMID: 31268708.

\bibitem{Campos-Gonzalez-Angulo:2019aa}
Jorge~A. Campos-Gonzalez-Angulo, Raphael~F. Ribeiro, and Joel Yuen-Zhou.
\newblock Resonant catalysis of thermally activated chemical reactions with
  vibrational polaritons.
\newblock {\em Nature Communications}, 10(1):4685, 2019.

\bibitem{Fontanesi2009}
L.~Fontanesi, L.~Mazza, and G.~C. La~Rocca.
\newblock Organic-based microcavities with vibronic progressions: Linear
  spectroscopy.
\newblock {\em Phys. Rev. B}, 80:235313, Dec 2009.

\bibitem{George2015-farad}
J.~George~{\it et al.}
\newblock Ultra-strong coupling of molecular materials: spectroscopy and
  dynamics.
\newblock {\em Faraday Discuss.}, 178:281--294, 2015.

\bibitem{Wu2016}
Ning Wu, Johannes Feist, and Francisco~J. Garcia-Vidal.
\newblock When polarons meet polaritons: Exciton-vibration interactions in
  organic molecules strongly coupled to confined light fields.
\newblock {\em Phys. Rev. B}, 94:195409, Nov 2016.

\bibitem{Zeb2018}
M.~Ahsan Zeb, Peter~G. Kirton, and Jonathan Keeling.
\newblock Exact states and spectra of vibrationally dressed polaritons.
\newblock {\em ACS Photonics}, 5(1):249--257, 2018.

\bibitem{Martinez-Martinez2018a}
Luis~A. Mart{\'{i}}nez-Mart{\'{i}}nez, Raphael~F. Ribeiro, Jorge
  Campos-Gonz{\'{a}}lez-Angulo, and Joel Yuen-Zhou.
\newblock {Can Ultrastrong Coupling Change Ground-State Chemical Reactions?}
\newblock {\em ACS Photonics}, 5(1):167--176, 2018.

\bibitem{kowalewski2016cavity}
Markus Kowalewski, Kochise Bennett, and Shaul Mukamel.
\newblock Cavity femtochemistry; manipulating nonadiabatic dynamics at avoided
  crossings.
\newblock {\em The journal of physical chemistry letters}, 2016.

\bibitem{Schafer2018}
Christian Sch{\"{a}}fer, Michael Ruggenthaler, and Angel Rubio.
\newblock {Ab initio nonrelativistic quantum electrodynamics: Bridging quantum
  chemistry and quantum optics from weak to strong coupling}.
\newblock {\em Physical Review A}, 98(4), 2018.

\bibitem{Groenhof2019}
Gerrit Groenhof, Cl{\`a}udia Climent, Johannes Feist, Dmitry Morozov, and
  J.~Jussi Toppari.
\newblock Tracking polariton relaxation with multiscale molecular dynamics
  simulations.
\newblock {\em The Journal of Physical Chemistry Letters}, 10(18):5476--5483,
  2019.
\newblock PMID: 31453696.

\bibitem{Knoll2001}
Ludwig Kn{\"o}ll, Stefan Scheel, and Dirk-Gunnar Welsch.
\newblock Qed in dispersing and absorbing media.
\newblock In Jan Perina, editor, {\em Coherence and Statistics of Photons and
  Atoms}, Wiley Series in Lasers and Applications. Wiley-VCH, 2001.

\bibitem{Raabe2007}
Christian Raabe, Stefan Scheel, and Dirk-Gunnar Welsch.
\newblock Unified approach to qed in arbitrary linear media.
\newblock {\em Phys. Rev. A}, 75:053813, May 2007.

\bibitem{Kena-Cohen2010}
S.~Kena-Cohen and S.~R. Forrest.
\newblock Room-temperature polariton lasing in an organic single-crystal
  microcavity.
\newblock {\em Nat Photon}, 4(6):371--375, 06 2010.

\bibitem{Wersall2019}
Martin Wers{\"a}ll, Battulga Munkhbat, Denis~G. Baranov, Felipe Herrera,
  Jianshu Cao, Tomasz~J. Antosiewicz, and Timur Shegai.
\newblock Correlative dark-field and photoluminescence spectroscopy of
  individual plasmon--molecule hybrid nanostructures in a strong coupling
  regime.
\newblock {\em ACS Photonics}, 6(10):2570--2576, 10 2019.

\bibitem{Climent2019}
Cl{\`a}udia Climent, Javier Galego, Francisco~J. Garcia-Vidal, and Johannes
  Feist.
\newblock Plasmonic nanocavities enable self-induced electrostatic catalysis.
\newblock {\em Angewandte Chemie International Edition}, 58(26):8698--8702,
  2019.

\bibitem{Bahsoun2018nanofluidics}
Hadi Bahsoun, Thibault Chervy, Anoop Thomas, Karl B{\"o}rjesson, Manuel
  Hertzog, Jino George, Elo{\"\i}se Devaux, Cyriaque Genet, James~A. Hutchison,
  and Thomas~W. Ebbesen.
\newblock Electronic light--matter strong coupling in nanofluidic
  fabry--p{\'e}rot cavities.
\newblock {\em ACS Photonics}, 5(1):225--232, 2018.

\bibitem{PerezRios2017}
Jes{\'{u}}s P{\'{e}}rez-R{\'{\i}}os, May~E Kim, and Chen-Lung Hung.
\newblock Ultracold molecule assembly with photonic crystals.
\newblock {\em New Journal of Physics}, 19(12):123035, dec 2017.

\bibitem{Ruddell:17}
S.~K. Ruddell, K.~E. Webb, I.~Herrera, A.~S. Parkins, and M.~D. Hoogerland.
\newblock Collective strong coupling of cold atoms to an all-fiber ring cavity.
\newblock {\em Optica}, 4(5):576--579, May 2017.

\bibitem{Fleming1986}
G.~Fleming.
\newblock {\em Chemical applications of ultrafast spectroscopy}.
\newblock Oxford University Press,New York, NY, United States, 1986.

\bibitem{Horng1997}
M.-L. Horng, J.~A. Gardecki, and M.~Maroncelli.
\newblock Rotational dynamics of coumarin 153 time-dependent friction,
  dielectric friction, and other nonhydrodynamic effects.
\newblock {\em The Journal of Physical Chemistry A}, 101(6):1030--1047, 1997.

\bibitem{Owrutsky1994}
J~C Owrutsky, D~Raftery, and R~M Hochstrasser.
\newblock Vibrational relaxation dynamics in solutions.
\newblock {\em Annual Review of Physical Chemistry}, 45(1):519--555, 1994.
\newblock PMID: 7811356.

\bibitem{Yi1996}
J.~Yi and J.~Jonas.
\newblock Raman study of vibrational and rotational relaxation of liquid
  benzene-d6 confined to nanoporous silica glasses.
\newblock {\em The Journal of Physical Chemistry}, 100(42):16789--16793, 01
  1996.

\bibitem{Coles2014}
DM~Coles, Niccolo Somaschi, Paolo Michetti, C.~Clark, Pavlos~G. Lagoudakis,
  P.~G. Savvidis, and David~G. Lidzey.
\newblock {Polariton-Mediated Energy Transfer between Organic Dyes in a
  Strongly Coupled Optical Microcavity}.
\newblock {\em Nature Mater.}, 13(May):712--719, 2014.

\bibitem{Tokmakoff1995}
A.~Tokmakoff, R.~S. Urdahl, D.~Zimdars, R.~S. Francis, A.~S. Kwok, and M.~D.
  Fayer.
\newblock Vibrational spectral diffusion and population dynamics in a
  glass‐forming liquid: Variable bandwidth picosecond infrared spectroscopy.
\newblock {\em The Journal of Chemical Physics}, 102(10):3919--3931, 1995.

\bibitem{Berden2000}
Giel Berden, Rudy Peeters, and Gerard Meijer.
\newblock Cavity ring-down spectroscopy: Experimental schemes and applications.
\newblock {\em International Reviews in Physical Chemistry}, 19(4):565--607,
  2000.

\bibitem{Haynes2005}
Christy~L. Haynes, Adam~D. McFarland, and Richard~P. Van~Duyne.
\newblock Surface-enhanced raman spectroscopy.
\newblock {\em Analytical Chemistry}, 77(17):338 A--346 A, 09 2005.

\bibitem{KUNDU2008}
Janardan Kundu, Fei Le, Peter Nordlander, and Naomi~J. Halas.
\newblock Surface enhanced infrared absorption (seira) spectroscopy on
  nanoshell aggregate substrates.
\newblock {\em Chemical Physics Letters}, 452(1):115 -- 119, 2008.

\bibitem{Jensen2000}
T.~R. Jensen, R.~P.~Van Duyne, S.~A. Johnson, and V.~A. Maroni.
\newblock Surface-enhanced infrared spectroscopy: A comparison of metal island
  films with discrete and nondiscrete surface plasmons.
\newblock {\em Applied Spectroscopy}, 54(3):371--377, 2000.

\bibitem{Rodriguez2007}
Kenneth~R. Rodriguez, Hong Tian, Joseph~M. Heer, Shannon Teeters-Kennedy, and
  James~V. Coe.
\newblock Interaction of an infrared surface plasmon with an excited molecular
  vibration.
\newblock {\em The Journal of Chemical Physics}, 126(15):151101, 2019/11/06
  2007.

\bibitem{Neubrech2012}
Frank Neubrech and Annemarie Pucci.
\newblock Plasmonic enhancement of vibrational excitations in the infrared.
\newblock {\em IEEE Journal of selected topics in quantum electronics},
  19(3):4600809--4600809, 2012.

\bibitem{Autore2018}
Marta Autore, Peining Li, Irene Dolado, Francisco~J Alfaro-Mozaz, Ruben
  Esteban, Ainhoa Atxabal, F{\`e}lix Casanova, Luis~E Hueso, Pablo
  Alonso-Gonz{\'a}lez, Javier Aizpurua, Alexey~Y Nikitin, Sa{\"u}l V{\'e}lez,
  and Rainer Hillenbrand.
\newblock Boron nitride nanoresonators for phonon-enhanced molecular
  vibrational spectroscopy at the strong coupling limit.
\newblock {\em Light: Science \& Applications}, 7(4):17172--17172, 2018.

\end{thebibliography}

\end{document}